\documentstyle[epsfig]{mn}
\onecolumn

\input epsf
\input rotate

\newif\ifAMStwofonts

\ifoldfss

  \ifCUPmtlplainloaded \else

    \NewTextAlphabet{textbfit} {cmbxti10} {}

    \NewTextAlphabet{textbfss} {cmssbx10} {}

    \NewMathAlphabet{mathbfit} {cmbxti10} {} 

    \NewMathAlphabet{mathbfss} {cmssbx10} {} 

  \fi

  \ifAMStwofonts

    \ifCUPmtlplainloaded \else

      \NewSymbolFont{upmath} {eurm10}

      \NewSymbolFont{AMSa} {msam10}

      \NewMathSymbol{\upi}     {0}{upmath}{19}

      \NewMathSymbol{\umu}     {0}{upmath}{16}

      \NewMathSymbol{\upartial}{0}{upmath}{40}

      \NewMathSymbol{\leqslant}{3}{AMSa}{36}

      \NewMathSymbol{\geqslant}{3}{AMSa}{3E}

      \let\leq=\leqslant \let\le=\leqslant

      \let\geq=\geqslant \let\ge=\geqslant

    \fi

  \fi

\fi 

\ifnfssone

  \newmathalphabet{\mathit}

  \addtoversion{normal}{\mathit}{cmr}{m}{it}

  \addtoversion{bold}{\mathit}{cmr}{bx}{it}

  \newmathalphabet{\mathbfit} 

  \addtoversion{normal}{\mathbfit}{cmr}{bx}{it}

  \addtoversion{bold}{\mathbfit}{cmr}{bx}{it}

  \newmathalphabet{\mathbfss} 

  \addtoversion{normal}{\mathbfss}{cmss}{bx}{n}

  \addtoversion{bold}{\mathbfss}{cmss}{bx}{n}

  \ifAMStwofonts

    \ifCUPmtlplainloaded \else

      \UseAMStwoboldmath

      \makeatletter

      \new@mathgroup\upmath@group

      \define@mathgroup\mv@normal\upmath@group{eur}{m}{n}

      \define@mathgroup\mv@bold\upmath@group{eur}{b}{n}

      \edef\UPM{\hexnumber\upmath@group}

      \new@mathgroup\amsa@group

      \define@mathgroup\mv@normal\amsa@group{msa}{m}{n}

      \define@mathgroup\mv@bold\amsa@group{msa}{m}{n}

      \edef\AMSa{\hexnumber\amsa@group}

      \makeatother

      \mathchardef\upi="0\UPM19

      \mathchardef\umu="0\UPM16

      \mathchardef\upartial="0\UPM40

      \mathchardef\leqslant="3\AMSa36

      \mathchardef\geqslant="3\AMSa3E

      \let\leq=\leqslant \let\le=\leqslant

      \let\geq=\geqslant \let\ge=\geqslant

    \fi

  \fi

\fi 

\ifnfsstwo

  \DeclareMathAlphabet{\mathbfit}{OT1}{cmr}{bx}{it}

  \SetMathAlphabet\mathbfit{bold}{OT1}{cmr}{bx}{it}

  \DeclareMathAlphabet{\mathbfss}{OT1}{cmss}{bx}{n}

  \SetMathAlphabet\mathbfss{bold}{OT1}{cmss}{bx}{n}

  \ifAMStwofonts

    \ifCUPmtlplainloaded \else

      \DeclareSymbolFont{UPM}{U}{eur}{m}{n}

      \SetSymbolFont{UPM}{bold}{U}{eur}{b}{n}

      \DeclareSymbolFont{AMSa}{U}{msa}{m}{n}

      \DeclareMathSymbol{\upi}{0}{UPM}{"19}

      \DeclareMathSymbol{\umu}{0}{UPM}{"16}

      \DeclareMathSymbol{\upartial}{0}{UPM}{"40}

      \DeclareMathSymbol{\leqslant}{3}{AMSa}{"36}

      \DeclareMathSymbol{\geqslant}{3}{AMSa}{"3E}

      \let\leq=\leqslant \let\le=\leqslant

      \let\geq=\geqslant \let\ge=\geqslant

    \fi

  \fi

\fi 

\ifCUPmtlplainloaded \else

  \ifAMStwofonts \else 

    \def\upi{\pi}

    \def\umu{\mu}

    \def\upartial{\partial}

  \fi

\fi

\begin{document}

\title{Three-integral models for axisymmetric galactic discs }

\author[B.Famaey, K.Van Caelenberg \& H.Dejonghe]{B. Famaey$^{1}$, 
K. Van Caelenberg$^{2}$ and H. Dejonghe$^{2}$\\
$^{1}$Institut d'Astronomie et d'Astrophysique CP226, Universit\'e Libre de Bruxelles, Boulevard du Triomphe, B-1050 Bruxelles, Belgium.\\
 Ph.D. student F.R.I.A., E-mail: bfamaey@astro.ulb.ac.be \\
$^{2}$Sterrenkundig Observatorium, Universiteit Gent, Krijgslaan 281, 
B-9000 Gent, Belgium
}

\date{Accepted ...
      Received 2001 december}

\pagerange{\pageref{firstpage}--\pageref{lastpage}}

\pubyear{2001}

\maketitle

\label{firstpage}

\begin{abstract}
We present new equilibrium component distribution functions that depend on three analytic integrals in a St\"ackel potential, and that can be used to model stellar discs of galaxies. These components are generalizations of two-integral ones and can thus provide thin discs in the two-integral approximation. Their most important properties are the partly analytical expression for their moments, the disc-like features of their configuration space densities (exponential decline in the galactic plane and finite extent in the vertical direction) and the anisotropy of their velocity dispersions. We further show that a linear combination of such components can fit a van der Kruit disc.

\end{abstract}


\section{Introduction}
It has been known for a long time that the classical two-integral
equilibrium theory in axisymmetric geometry is not sufficient to
adequately describe the stellar discs of galaxies. In accordance with Jeans theorem (Jeans 1915), the phase space distribution function of a stellar system in a steady state depends only on the isolating integrals of the motion; the binding energy $E$ and the vertical component of the angular momentum $L_z$ are isolating integrals in a stationary and axisymmetric system.
It is a fundamental property of all two-integral
distribution functions $F(E,L_z)$ that the dispersion of the velocity
in the radial direction equals the dispersion in
the vertical direction: we know that, for example, the disc of the Milky
Way does not have that property\ (Binney \& Merrifield 1998).
Illustrations of other shortcomings of a two-integral model in a galactic
context can be found in Durand, Dejonghe \& Acker\ (1996).

The introduction of a third integral of the motion helps to overcome these 
constraints: in that case, the velocity dispersions can be different in all 
the directions.  Numerical experiments show that 
a third isolating integral seems to exist
for most orbits in realistic galactic potentials\ (Ollongren 1962, Innanen \& 
Papp 1977, Richstone 1982). This third integral can be
taken into account numerically in the models by using extensions of 
Schwarzschild's\ (1979)
orbit superposition technique\ (Cretton et al. 1999, Zhao 1999, H\"afner et al. 2000). It is also possible
to define an analytic third integral specific to particular orbital families\ (de Zeeuw, Evans \& Schwarzschild 1996, Evans, H\"afner \& de Zeeuw 1997) or an approximate global third integral\ (Petrou 1983ab, Dehnen \& Gerhard 1993), but
we choose to construct models with an exact analytic third integral by using a St\"ackel potential\ (St\"ackel
1890, de Zeeuw 1985). 

It is not quite obvious to define suitable global distribution
functions $F(E,L_z,I_3)$ that depend on three exact analytic integrals and that can
somewhat realistically represent our ideas of a real stellar disc.
For example, Bienaym\'e\ (1999) made three-integral extensions of
the two-integral parametric distribution functions described in Bienaym\'e \& 
S\'echaud\ (1997), but these ones were built to model the kinematics of
neighbouring stars in the Milky Way only.
Dejonghe \& Laurent\ (1991) also defined the three-integral Abel distribution
functions, but these ones could not provide very thin discs in the 
two-integral approximation. Robijn \& de Zeeuw\ (1996) constructed three-integral distribution functions
for oblate galaxy models, but they also had problems to recover the two-integral approximation.

In this paper we continue the work of Batsleer \& Dejonghe\ (1995), who 
constructed component distribution functions that are two-integral, but that 
can represent (very) thin discs when a judicious linear combination of them is 
chosen. We use these components as a basis for new component distribution
functions that are three-integral, of which the Batsleer \& Dejonghe 
components are a special case.

In the next section, we outline some fundamentals of two-integral equilibrium systems and we show how to model discs with a finite extent in the vertical direction. In section 3, we present some general facts about St\"ackel potentials and we present new analytic three-integral distribution functions that can represent stellar discs. An analytical expression for the moments of these distribution functions is calculated in section 4. In the next section, we discuss their physical properties and show their realistic disc-like character. Finally, in section 6, we show that these distribution functions  can be used as basis functions in the modeling of a van der Kruit disc. For the conclusions, we refer to section 7.

\section{two-integral fundamentals}

We  denote the gravitational potential in cylindrical coordinates 
$(\varpi,\phi,z)$ by $V(\varpi,z)=-\psi(\varpi,z)\le0$. The 
two classical isolating integrals of the
motion are the binding energy, $E=\psi(\varpi,z)-\frac{1}{2}v^2$, and the 
$z$-component of the angular momentum, $L_z=\varpi v_{\phi}=\varpi^2\dot\phi$.

When we use the term orbit, we do not consider the information
contained in the phases of the orbital motion: an orbit is thus
shorthand for an orbital density. Even with this definition, each pair
$(E,L_z)$ represents a family of orbits; to uniquely identify one
particular orbit, we need the presence of a third effective integral
of the motion (i.e.\ we need a triple $(E,L_z,I_3)$). 

The axisymmetric nature of the system implies we can focus on the motion
in a meridional plane (i.e.\ a plane with constant $\phi$). For a position 
$(\varpi_0,z_0)$ in the meridional plane, the
expressions for $E$ and $L_z$ imply that all families of
orbits that visit this position have isolating integrals of the motion
$(E,L_z)$ that meet the requirement
\begin{equation}
        E \le \psi(\varpi_0,z_0) - \frac{L_z^2}{2\varpi_0^2},
\label{eq:Ele}
\end{equation}
since we know that
\begin{equation}
        v_{\varpi}^2 + v_z^2 \ge 0.
\label{eq:Mv}
\end{equation}
For given $E$ and $L_z$, this relation restricts possible motion for
the corresponding family of orbits to a toroidal volume in
configuration space.

In $(E,L_z^2)$-space, Eq.\ (\ref{eq:Ele})
defines the region in which the points correspond to families of
orbits passing through $(\varpi_0,z_0)$. The boundary line (equality
in Eq.\ (\ref{eq:Ele})) contains orbits that reach the given position
with zero velocity (\ref{eq:Mv}) in the meridional plane. Keeping
$z=z_0$ fixed while allowing $\varpi$ to vary then gives us a family
of such boundary lines, of which we denote the envelope by
\begin{equation}
        E = S_{z_0}(L_z),
\end{equation}
with the parametric equations\ (Batsleer \& Dejonghe 1995, Eq. 4 \& 5)
\begin{eqnarray}
        \left\{ \begin{array}{lll}
                E & = & \displaystyle\psi(\varpi,z_0) - \frac{L_z^2}{2\varpi^2} \\[1ex]
                L_z^2 & = \displaystyle & - \varpi^3 \frac{\partial \psi}{\partial \varpi}
                        (\varpi,z_0).
                \end{array}
        \right.
\end{eqnarray}
All points in integral space with $E < S_{z_0}(L_z)$ represent
families of orbits that will pass through $z = z_0$ at a certain
$\varpi$. Orbits for which $E = S_{z_0}(L_z)$ also do reach the height
$z_0$, but can never go any higher. All points in integral space with
$E > S_{z_0}(L_z)$ represent families of orbits that cannot reach $z
= z_0$. $S_{z_0}$ is thus the minimal binding energy of an orbit that cannot bring a star higher than $z_0$ above the galactic plane.

For every height $z_1 > z_0$ we find a similar curve $E =
S_{z_1}(L_z)$, with $S_{z_1}(L_z) < S_{z_0}(L_z)$ for every value of
$L_z$. Similarly, the envelope for the orbits that cannot go higher
than $z=0$ is given by $E = S_{0}(L_z)$, which gives us all circular
orbits in the galactic plane.
 
Orbits belonging to a disc with a maximum height $z_0$ are thus given
by $(E,L_z)$ for which $S_{z_0}(L_z) \le E \le S_0(L_z)$ (the shaded
area in Figure 1). Batsleer \& Dejonghe\ (1995) constructed disc-like
component distribution functions with a finite extent in vertical
direction by setting them equal to zero for $E < S_{z_0}(L_z)$.
In order to fully understand the components that we present here,
that paper should probably be considered as preparatory reading.

\begin{figure}

\vbox{\epsfxsize=10cm \epsfbox{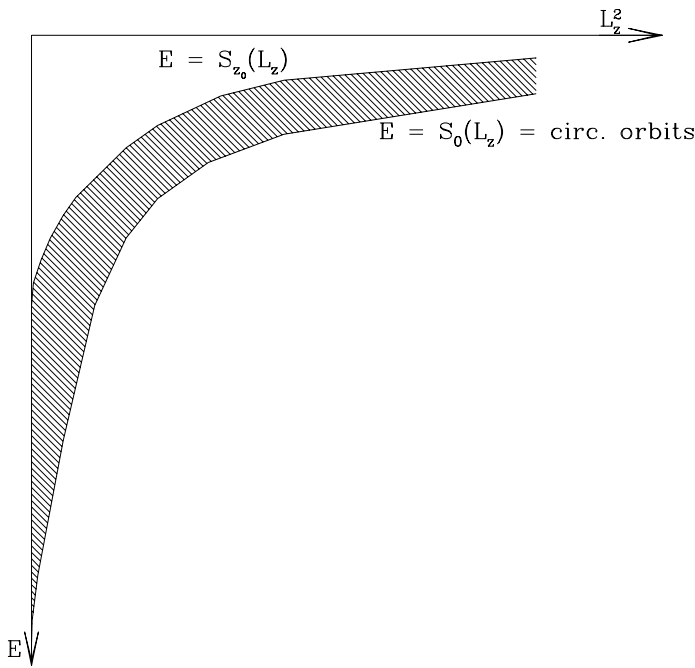}\vspace{-3cm}}
\hfill\parbox[b]{5.5cm}{\caption[]{The shaded area is the area in the $(E,L_z^2)$-plane for which the corresponding families of orbits cannot go higher than $z_0$ above the galactic plane. The distribution function of a stellar disc with maximum height $z_0$ is null outside of this area.}}  
  \label{fig:fig2int}
\end{figure}

\section{Construction of three-integral components}

\subsection{Spheroidal coordinates and St\"ackel potentials}

\noindent
We will work in {\it spheroidal coordinates}, since these coordinates
allow a simple expression for our (axisymmetric) St\"ackel potential. 
Spheroidal coordinates are given by $(\lambda,\phi,\nu)$,
with $\lambda$ and $\nu$ the roots for $\tau$ of
\begin{eqnarray}
        \frac{\varpi^2}{\tau+\alpha}+\frac{z^2}{\tau+\gamma}=1 & & \alpha < \gamma < 0,
\end{eqnarray}
and $(\varpi,\phi,z)$ cylindrical coordinates. The parameters $\alpha$
and $\gamma$ are both constant and smaller than zero.

A potential is of {\it St\"ackel form}, if there exists a spheroidal
coordinate system $(\lambda,\phi,\nu)$, in which the potential can be
written as
\begin{equation}
        V(\lambda,\nu) = - \frac{f(\lambda) - f(\nu)}{\lambda - \nu},
\end{equation}
for an arbitrary function $f(\tau) = (\tau+\gamma)G(\tau)$, $\tau =
\lambda,\nu$. The function $-G(\lambda)$ then represents the potential
in the $z=0$ plane.

For this kind of potential, the Hamilton-Jacobi equation is separable in spheroidal coordinates, and 
therefore the orbits admit three analytic isolating integrals of the motion.
The third integral of galactic dynamics has the form 
\begin{equation}
       I_3 = \frac{1}{2} (L_x^2 + L_y^2) + (\gamma - \alpha) {\left[\frac{1}{2} v_z^2 - z^2 \frac{G(\lambda) - G(\nu)}{\lambda - \nu}\right]}
\end{equation}

More details can be found in de Zeeuw\ (1985) and in Dejonghe \& de Zeeuw\ 
(1988).

\subsection{Modified Fricke components}

As mentioned before, we intend to create three-integral stellar
distribution functions, for the construction of stellar discs:
we want to achieve an exponential decline in the mass density for large
radii, while we want to introduce a preference for (almost) circular
orbits.

It has been known for some time that two-integral models can describe very thin disc systems, with the restriction that both vertical and radial dispersions are equal\ (Jarvis \& Freeman 1985). So we want to create three-integral distribution functions that can describe very thin discs in the two-integral approximation, unlike the Abel distribution functions\ (Dejonghe \& Laurent 1991).

The Fricke components\ (Fricke 1952) $E^{\eta}L_z^{\beta}$ favour that part of
phase space where stars populate circular orbits, so they could be taken as a 
starting point. However, they cannot be used in their basic form to model discs 
with a finite extent because they populate orbits which can reach arbitrary 
large heights: therefore, we will take as a starting point the components 
defined in Batsleer \& Dejonghe\ (1995, Eq. 19).

In order to make the components depending on the third integral $I_3$, we 
introduce the factor $(p + qE + rL_z^2 + s I_3)^\delta$ in which the parameter
$s$ (and $\delta$) will be responsible for the three-integral character of the
components. The coefficients $p$, $q$, $r$ and $s$ can, in the most general 
case, be functions of $L_z$.

This leads us towards a general three-integral disc component
of the form

\begin{eqnarray}
            F(E,L_z,I_3) = f(L_z) {\left(\frac{E - S_{z_0}(L_z)}{S_0(L_z) - S_{z_0}(L_z)} \right)}^{\eta}(p+qE+rL_z^2+sI_3)^\delta 
\label{eq:3icomp}
\end{eqnarray}
\noindent
if
\begin{equation}
                    \left\{ 
                        \begin{array}{l}
                            E - S_{z_0}(L_z) \geq 0 \\
                            p + qE + rL_z^2 + sI_3 \geq 0 \\ 
                        \end{array}
                    \right..
\label{eq:cond}
\end{equation}
The distribution function is identically zero in all other cases.
The function $f(L_z)$ is defined as
\begin{eqnarray}
        f(L_z)  =  \frac{1}{1+e^{-a L_z}} {(2 L_z^2 S_{z_0}(L_z))}^\beta 
                {(S_{z_0}(L_z))}^{\alpha_1} e^{-\frac{\alpha_2}{S_{z_0}(L_z)}}.
\end{eqnarray}

The parameter $a$ is the rotation parameter (the value of $a$ influences only
the odd moments of $F$, see section 4). If $a=0$, there is no rotation for the
component, and if $a=+\infty$, $F$ represents a {\it maximum streaming} 
component with no counter-rotating stars.

The requested exponential decline in the mass density with large radii
is controlled by the parameter $\alpha_2$ (and to some extent by the
parameter $\alpha_1$, see section 5).

The parameter $\eta$ is responsible for the favouring of almost circular orbits, i.e. 
orbits with a binding energy $E$ as close as possible to that of circular 
orbits in the galactic plane (see section 5).

Furthermore, if we want to favour (almost) circular orbits, we have to suppose 
the distribution function to be an increasing function of $E$: this forces $q$ 
to be positive. 

Since a large $I_3$ implies that the orbit can reach a large height
above the galactic plane\ (see de Zeeuw 1985 for a complete analysis of the 
orbits in a St\"ackel potential), the orbits with small $I_3$'s have to be 
favoured in order to describe thin discs: this forces $s$ to be negative. 

Other constraints (on $p$, $q$, $r$, $s$ and $\eta$)
will be imposed in section 4, in order to enable the analytical calculations of the moments.


\section{Moments}

\subsection{Theory}

The moments of a distribution function $F$ at the point $(\lambda,\phi,\nu)$ 
of a spheroidal coordinate system are defined as
\begin{equation}
        \mu_{l,m,n}(\lambda,\nu)= \int\! \int\! \int\! F(E,L_z,I_3) v_\lambda^l
               v_\phi^m v_\nu^n dv_\lambda dv_\phi dv_\nu,      
\label{defmoments}
\end{equation}
with $v_\lambda$, $v_\phi$, $v_\nu$ the components of the velocity
in the $\lambda$, the $\phi$ and the $\nu$ direction of the spheroidal
coordinate system, and $l$, $m$, $n$ integers.

\noindent
The mass density, the mean velocity and the velocity dispersions of the stellar
system represented by $F$ can easily be expressed in terms of the moments
(\ref{defmoments}) by
\begin{eqnarray}
        \lefteqn{\rho(\lambda,\nu) = \mu_{0,0,0}(\lambda,\nu) \nonumber} \\
        \lefteqn{\rho\langle v_{\phi} \rangle(\lambda,\nu) = \mu_{0,1,0}(\lambda,\nu) \nonumber} \\
        \lefteqn{\rho\sigma_{\lambda}^2(\lambda,\nu) = \mu_{2,0,0}(\lambda,\nu) \nonumber} \\
        \lefteqn{\rho\langle v_{\phi}^2 \rangle(\lambda,\nu) = \mu_{0,2,0}(\lambda,\nu) \nonumber} \\
        \lefteqn{\rho\sigma_{\nu}^2(\lambda,\nu) = \mu_{0,0,2}(\lambda,\nu)} 
\end{eqnarray}

To obtain the value of one of these moments, we have to integrate over the 
volume in velocity-space corresponding to all orbits that pass through the 
point $(\lambda,\phi,\nu)$.

\noindent
Since all three integrals of the motion are quadratic in $v_\lambda$ and 
$v_\nu$, if $l$ or $n$ is an odd integer, the moment $\mu_{l,m,n}$ is
identically zero.
If $l$ and $n$ are even integers, the moment $\mu_{l,m,n}$ can be written as an
integral computed in the integral space. We assume in sections 4.1 and 4.2 that 
$a=0$ (in that case, there is no rotation and if $m$ is odd, the moment is zero) and that $m$ is an even
integer (the general case will easily be derived from this one in section 4.3). Under these assumptions,
 
\begin{eqnarray}
        \mu_{l,m,n} & = & \frac{2^{\frac{l+n}{2}+ 1}}
        {\varpi (\lambda-\nu)^{\frac{l+n}{2}}} 
        \int\! \frac{f(L_z^2)}{\sqrt{L_z^2}} 
        v_\phi^m dL_z^2 \int\!\int\! dE  dI_3
        {\left( \frac{E - S_{z_0}}
           {S_0 - S_{z_0}} \right)}^{\eta} (p + qE + rL_z^2 + s I_3)^\delta 
                        (I_3^+ - I_3)^{\frac{l-1}{2}}
                        (I_3 - I_3^-)^{\frac{n-1}{2}} 
\label{eq:moment3icomp}
\end{eqnarray}

\noindent
where $I_3^+$ and $I_3^-$ are given by
\begin{equation}
        \displaystyle
        \left\{ \begin{array}{l}
                  I_3^+(E,L_z^2) = (\lambda+\gamma) [G(\lambda)-E] - 
                      \frac{\lambda+\gamma}{2(\lambda+\alpha)} L_z^2 \\[2mm]
                  I_3^-(E,L_z^2) = (\nu+\gamma) [G(\nu)-E] - 
                      \frac{\nu+\gamma}{2(\nu+\alpha)} L_z^2.
               \end{array}
        \right.
\end{equation}

\noindent
More details can be found in Dejonghe \& de Zeeuw\ (1988).

For our components given by Equation (\ref{eq:3icomp}), the integration surface
in the $(E,I_3)$-plane is defined by
\begin{equation}
        \displaystyle
        \left\{ \begin{array}{l}
                I_3^-(E,L_z^2) \leq I_3 \leq I_3^+(E,L_z^2)\\[2mm]
                S_{z_0}(L_z) \leq E \leq \psi-\frac{L_z^2}{2\varpi^2}\\[2mm]
                p + qE + rL_z^2 + s I_3  \geq  0.
                \end{array}
        \right.
\end{equation}

\noindent
The integration limits for the integral in $L_z^2$ will be determined in 
section 4.3.

\subsection{(Partly) Analytical expression for the moments}

We want to reduce the triple integral (13) to a
simple one by solving the innermost double integral analytically. 
The reader not interested in the mathematical details might step directly to section 5.

\begin{figure} 
\vspace{0cm}
\hbox{\hspace{0cm}\epsfxsize=8.8cm \epsfbox{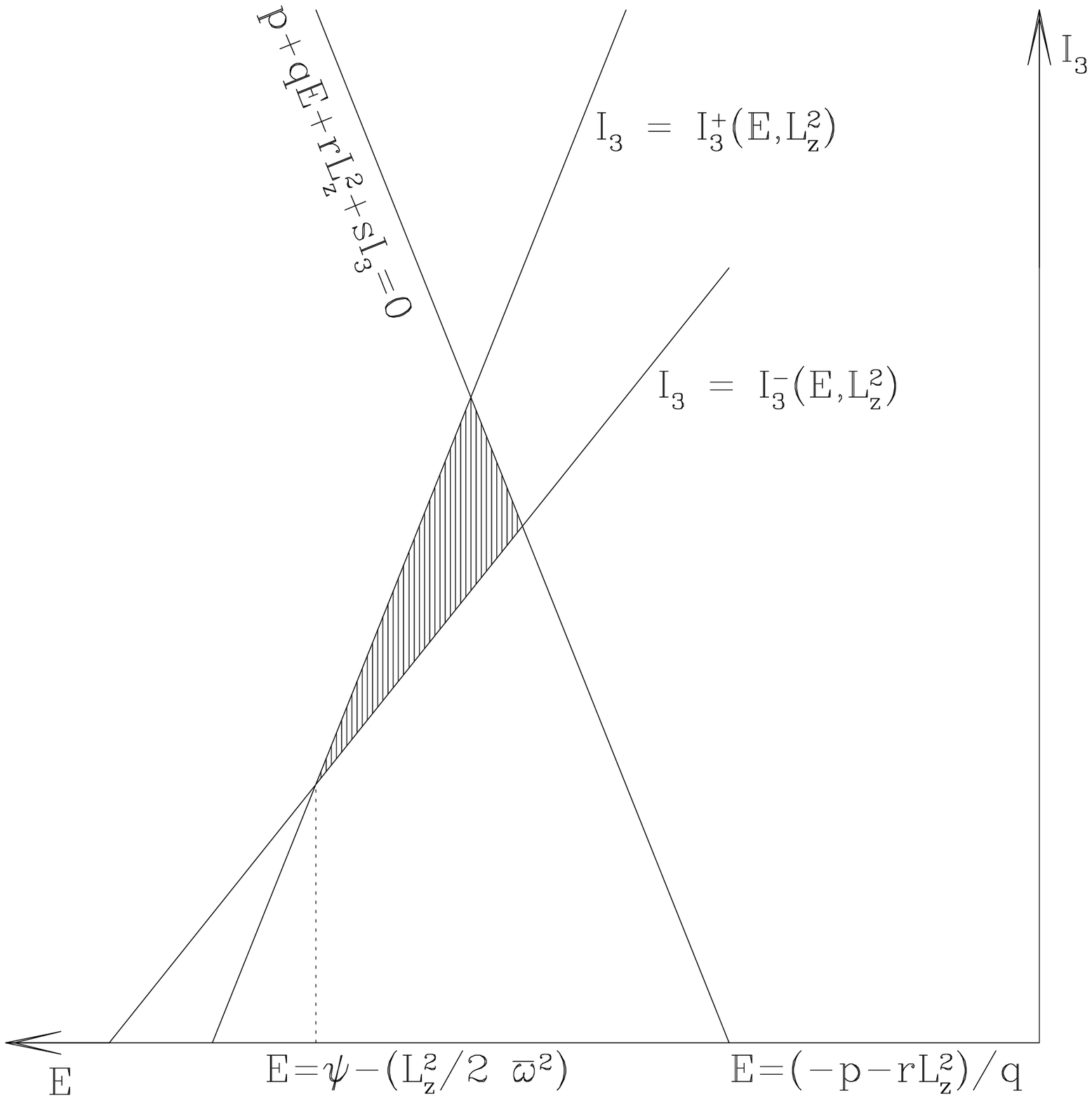}
\epsfxsize=8.8cm \epsfbox{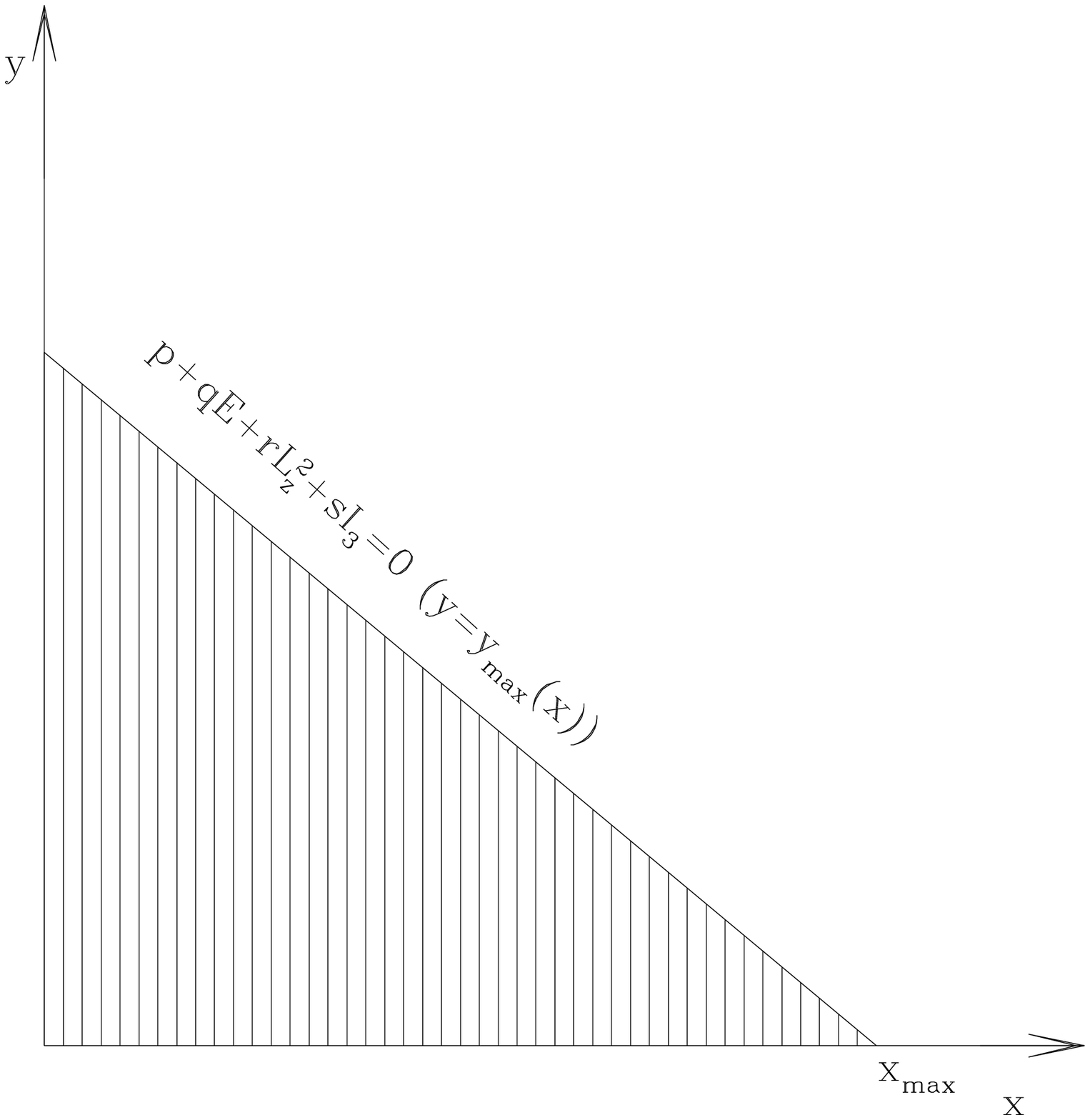}}
\vspace{0cm}
  \caption{The integration area for the inner double integral in Equation (13) in the $(E,I_3)$-plane (left panel, $L_z$ is fixed) and in the $(x(E,I_3),y(E,I_3))$-plane (right panel, see Equation 19).}
  \label{fig:integration_areas}
\end{figure}

For this double integral to be analytically solved, however, we will have to
make use of those combinations of $p$, $q$, $r$ and $s$ (see section 4.3)
for which the integration area is transformed into the triangle bounded by 
\begin{equation}
        \displaystyle
        \left\{ \begin{array}{l}
                I_3  =  I_3^+(E,L_z^2) \\
                I_3  =  I_3^-(E,L_z^2) \\
                p + qE + rL_z^2 + s I_3  =  0,
               \end{array}
        \right.
\end{equation}
as shown in Figure \ref{fig:integration_areas}. In this situation, we
can express the factor $(E - S_{z_0})/ (S_0 - S_{z_0})$ as a linear
combination of the other three factors in the integrandum (corresponding to the
bounding lines of the integration surface):
\begin{eqnarray}
        \frac{E - S_{z_0}}{S_0 - S_{z_0}} &=&   
           t (I_3^+ - I_3) + u (I_3 - I_3^-) + 
        v (p + qE + rL_z^2 + s I_3).
\label{eq:summation}
\end{eqnarray}
We impose $\eta$ to be  an integer: then the double integral in the $(E,I_3)$-plane 
transforms into a sum of integrals:
\begin{eqnarray}
        \lefteqn{\sum_{i=0}^\eta  \; \sum_{j=0}^{\eta-i} 
                \left( \! \begin{array}{c}
                          \eta \\ i
                       \end{array}
                \! \right)
                \left( \! \begin{array}{c}
                          \eta - i \\ j
                       \end{array}
                \! \right) 
            v^i \; t^{\eta-i-j} \; u^j  \int\! \int\! dE \,dI_3 \; 
            (p + qE + rL_z^2 + s I_3)^{\delta+i} \; (I_3^+ - I_3)^{\eta+\frac{l-1}{2}-i-j} 
           \; (I_3 - I_3^-)^{\frac{n-1}{2}+j}}             
\label{eq:dedi3sum}
\end{eqnarray}
For each integral in this summation, the integrandum consists of
factors whose zero-points define the bounding lines for the
integration surface in the $(E,I_3)$-plane. These integrals can be
solved analytically.

We will be in this situation (for all the points $(\lambda,\phi,\nu)$ of configuration space, where the moments are calculated) whenever $p + qE + rL_z^2 + s I_3 = 0 $
does intersect the $E$-axis for $E \geq S_{z_0}(L_z)$.

In order to solve the integrals analytically, one uses
the new integration variables $x$ and $y$, defined by (for a fixed $L_z^2$)
\begin{equation}
        \displaystyle
        \left\{ 
        \begin{array}{l}
        x(E,I_3) = I_3^+(E) - I_3 \\
        y(E,I_3) = I_3 - I_3^-(E).
        \end{array}
        \right.
\end{equation}
The line in the $(E,I_3)$-plane $p + qE + rL_z^2 + s I_3 = 0 $ becomes 
$y = y_{\mathrm{ max}}(x)$, and the root
of $y_{\mathrm{ max}}(x) = 0$ is $x_{\mathrm{ max}}$ (see Figure \ref{fig:integration_areas}).

\noindent 
To make a more compact notation possible, we first define the auxiliary
functions $g(\tau)$ and $h(\tau)$, as
\begin{equation}
        g(\tau) = G(\tau) - \frac{L_z^2}{2(\tau+\alpha)}
\end{equation}
and
\begin{equation}
        h(\tau) = s(\tau+\gamma)-q.
\end{equation}

\noindent
We then have
\begin{eqnarray}
        \lefteqn{x_{\mathrm{ max}}
                 = -  \left\{ p + rL_z^2 - h(\nu)\psi(\lambda,\nu) + s(\nu+\gamma)G(\nu)
                   - \frac{L_z^2}{2\varpi^2} (q-sz^2)
           \right\} 
            \frac{(\lambda-\nu)}{h(\nu)}} \\
            & \equiv & - \frac{(\lambda-\nu)}{h(\nu)}  x_{\mathrm{ max}}'
\end{eqnarray}
and
\begin{equation}
        y_{\mathrm{ max}}(x)  =  \frac{h(\nu)}{h(\lambda)} (x_{\mathrm{ max}} - x)
\end{equation}

Solving the integral part of one of the terms in Eq.\
(\ref{eq:dedi3sum}) for $y$ then yields
\begin{eqnarray}
        \lefteqn{\frac{1}{(\lambda-\nu)^{\delta+i+1}}
            \int_0^{x_{\mathrm{ max}}}  x^{\eta +\frac{l-1}{2}-i-j} dx
            \int_0^{y_{\mathrm{ max}}(x)} y^{\frac{n-1}{2}+j}
            \left[ h(\lambda) (y-y_{\mathrm{ max}}(x)) \right]^{\delta+i} dy \hspace{.5cm} \nonumber}\\
         \lefteqn{=\frac{(-h(\lambda))^{\delta+i}}{(\lambda-\nu)^{\delta+i+1}}
             B(\frac{n-1}{2}+j+1,\delta+i+1)  
           \int_0^{x_{\mathrm{ max}}} x^{\eta +\frac{l-1}{2}-i-j}
            y_{\mathrm{ max}}(x)^{\frac{n-1}{2}+i+j+\delta+1} dx}
\label{eq:dedi3term}
\end{eqnarray}

After solving for $x$ (analogous to what we did for $y$), one obtains
for the whole summation (\ref{eq:dedi3sum})

\begin{eqnarray}
        \lefteqn{\sum_{i=0}^\eta \; \sum_{j=0}^{\eta-i} 
                \left( \! \begin{array}{c}
                          \eta \\ i
                       \end{array}
                \! \right)
                \left( \! \begin{array}{c}
                          \eta - i \\ j
                       \end{array}
                \! \right) 
            v^i \; t^{\eta-i-j}\; u^j   
        \frac{(-h(\nu))^{\delta+\frac{n+1}{2}+i+j}}{(\lambda-\nu)^{\delta+i+1}
          (-h(\lambda))^{\frac{n+1}{2}+j}}
        (x_{\mathrm{ max}})^{\delta+\eta+\frac{l+n}{2}+1} \nonumber} \\
        & & \times \frac{\Gamma(\eta+\frac{l+1}{2}-i-j) \; \Gamma(\frac{n+1}{2}+j)
                 \;  \Gamma(\delta+i+1)}
                {\Gamma(\delta+\eta+\frac{l+n}{2}+2)}
\end{eqnarray}
\noindent
The coefficients $t$, $u$ and $v$ are calculated by equalizing term by term
in equation (\ref{eq:summation})
\begin{equation}
        \displaystyle
        \left\{ \begin{array}{l}
                \frac{1}{S_0-S_{z_0}}  =  -t(\lambda+\gamma)+u(\nu+\gamma)+vq \\[2mm]
                0  =  -t + u + vs \\[2mm]
                -\frac{S_{z_0}}{S_0-S_{z_0}}  =  t(\lambda+\gamma){\left[G(\lambda)-\frac{L_z^2}{2(\lambda+\alpha)}\right]}-u(\nu+\gamma) 
                {\left[G(\nu)-\frac{L_z^2}{2(\nu+\alpha)}\right]}+vrL_z^2+vp
               \end{array}
        \right.
\end{equation}

\noindent
We find for the coefficients 
\begin{equation}
        v  =  \frac{v'}{x_{\mathrm{ max}}'}, u  =  \frac{u'}{(\lambda-\nu) x_{\mathrm{ max}}'}, t = \frac{t'}{(\lambda-\nu) x_{\mathrm{ max}}'}
\end{equation}

\noindent
with
\begin{equation}
        v'  =  \frac{\psi(\lambda,\nu)-S_{z_0}-\frac{L_z^2}{2\varpi^2}}{S_0-S_{z_0}}, 
        u'  =  -h(\lambda)v' - \frac{x_{\mathrm{ max}}'}{S_0 - S_{z_0}}, 
        t'  =  u' + s(\lambda-\nu)v'
\end{equation}

\noindent
Now we have a one-dimensional numerical integration to perform, like for the 
Abel components\ (Dejonghe \& Laurent 1991).

\subsection{The one-dimensional numerical integration}

The triple integral (13) is reduced to a simple one if we judiciously choose 
the parameters $p$, $q$, $r$ and $s$.
The double integral in the $(E,I_3)$-plane (for a fixed $L_z$) can be 
solved analytically whenever $p+qE+rL_z^2+sI_3 = 0$ does intersect the $E$-axis
for $E \geq S_{z_0}(L_z)$. It has to be the case for all the $L_z$ relevant in the integration. 
So if we take $p=-S_{z_0}$, $q=1$ and $r \leq 0$, the double integral can be solved
analytically because $E-S_{z_0}+rL_z^2+sI_3 = 0$ intersects the $E$-axis for
the value $E=S_{z_0}-rL_z^2 \geq S_{z_0}$.
In that case, the double condition (\ref{eq:cond}) becomes a simple one
because the condition $E-S_{z_0} \geq 0$ is automatically verified when
$E-S_{z_0}+rL_z^2+sI_3 \geq 0$ since $s \leq 0$ and $I_3 \geq 0$.

We now have to determine the integration limits of the simple integral in $L_z$. Since conditions (\ref{eq:Ele}) and (\ref{eq:cond}) imply that
\begin{equation}
       S_{z_0}(L_z) \leq E \leq \psi-\frac{L_z^2}{2\varpi^2},
\end{equation}
the integration limits for the integral in $L_z^2$ are, in a first time, 
determined by the intersections of $S_{z_0}(L_z)$ and the line
$E=\psi-\frac{L_z^2}{2\varpi^2}$ in the $(E,L_z^2)$-plane (see Figure 3).

\noindent
Furthermore, in order to have a non-degenerate (i.e. not empty) triangle in 
Figure \ref{fig:integration_areas}, we must have
\begin{equation}
x_{\mathrm{ max}} \geq 0   \Leftrightarrow   x_{\mathrm{ max}}' \geq 0.
\label{eq:xmax}
\end{equation}
Since the condition $\psi-\frac{L_z^2}{2\varpi^2}-S_{z_0}
\geq 0$ is automatically verified when condition (\ref{eq:xmax}) is 
verified, the equality in (\ref{eq:xmax}) fixes the minimal and maximal angular momentum to take into account in the integration.

Knowing that $L_z^2 = \varpi^2 v_{\phi}^2$, the resulting expression for 
the moment (with $a=0$ and $m$ even) becomes
\begin{eqnarray}
        \lefteqn{\mu_{l,m,n}  =  2^{\frac{l+n}{2}+\beta+1} \varpi^{2\beta}
                \frac{\Gamma(\delta+1)\Gamma(\frac{n+1}{2})
                        \Gamma(\eta+\frac{l+1}{2})}
                     {\Gamma(\delta+\eta+\frac{l+n}{2}+2)}
        \int_{(v_\phi^2)_{\mathrm{min}}}^{(v_\phi^2)_{\mathrm{max}}} (v_\phi^2)^{\beta+\frac{m-1}{2}} 
                (S_{z_0})^{\beta+\alpha_1}
                \frac{e^{-\frac{\alpha_2}{S_{z_0}}}}{2} 
                \nonumber}\\
        & & \times \frac{(x_{\mathrm{ max}}')^{\delta+\frac{l+n}{2}+1}}
                 {(-h(\lambda))^{\frac{n+1}{2}}(-h(\nu))^{\eta+\frac{l+1}{2}}}
                 \sum_{i=0}^\eta \; \sum_{j=0}^{\eta-i} 
                \left( \! \begin{array}{c}
                          \eta \\ i
                       \end{array}
                \! \right)
                \left( \! \begin{array}{c}
                          \eta - i \\ j
                       \end{array}
                \! \right) 
            v'^i \nonumber \\
        & & \times t'^{ \eta-i-j}\; u'^j 
            \frac{\Gamma(\delta+i+1)}{\Gamma(\delta+1)} \; \frac{\Gamma(\eta+\frac{l+1}{2}-i-j) \; \Gamma(\frac{n+1}{2}+j)}
                {\Gamma(\eta+\frac{l+1}{2}) \; \Gamma(\frac{n+1}{2})} 
        \; \frac{(-h(\nu))^{i+j}}{(-h(\lambda))^j} dv_\phi^2
\label{eq:final}
\end{eqnarray}

\noindent
This integration has to be performed numerically.

In the general case $a \not= 0$, the expression (\ref{eq:final}) is still
valid for the even moments. When $m$ is odd, the integrandum has to be multiplied by a factor $A$:
\begin{equation}
A = \frac{1-e^{-a \varpi |v_{\phi}|}}{1+e^{-a \varpi |v_{\phi}|}}
\label{eq:rotationfactor}
\end{equation}

\begin{figure}

\vbox{\epsfxsize=8cm \epsfbox{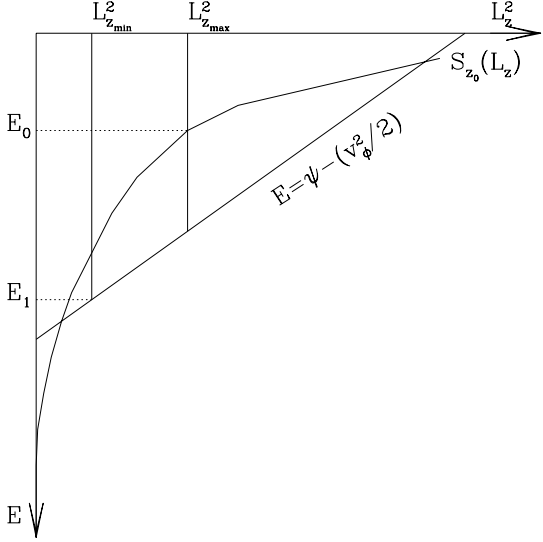}\vspace{-3.5cm}}
\hfill\parbox[b]{5.5cm}{\caption[]{The $(E,L_z^2)$-plane. The integrations limits in $L_z^2$, $(L_z)_{min}^2$ and $(L_z)_{max}^2$, must be between the intersections of the curve $E=S_{z_0}(L_z)$ and the line $E=\psi-\frac{v_{\phi}^2}{2}$. Condition (\ref{eq:xmax}) then definitively fixes these limits. $E_0$ and $E_1$ are the minimal and maximal energy taken into account in the integration.}}  
  \label{fig:intareaLz}
\end{figure}


\section{Physical properties of the components}

In this section, we show the realistic disc-like character of our
stellar distribution functions: their mass density has a finite extent
in the vertical direction and an exponential decline in the galactic
plane, they favour almost circular orbits and their velocity
dispersions are different in the vertical and radial direction. By
varying the parameters, we can give a wide range of shapes to the
components.

In order to illustrate the role of the different parameters, we
calculate the moments of many component distribution functions with
different values for the parameters.

As galactic potential, we use one of the St\"ackel potentials
described by Famaey \& Dejonghe\ (2001), that are extensions of the ones
described by Batsleer \& Dejonghe\ (1994).  In the implementation of
the theory we choose $p=-S_{z_0}$, $q=1$, and $r=0$.

\subsection{The parameter $z_0$}

\begin{figure}
\psfig{figure=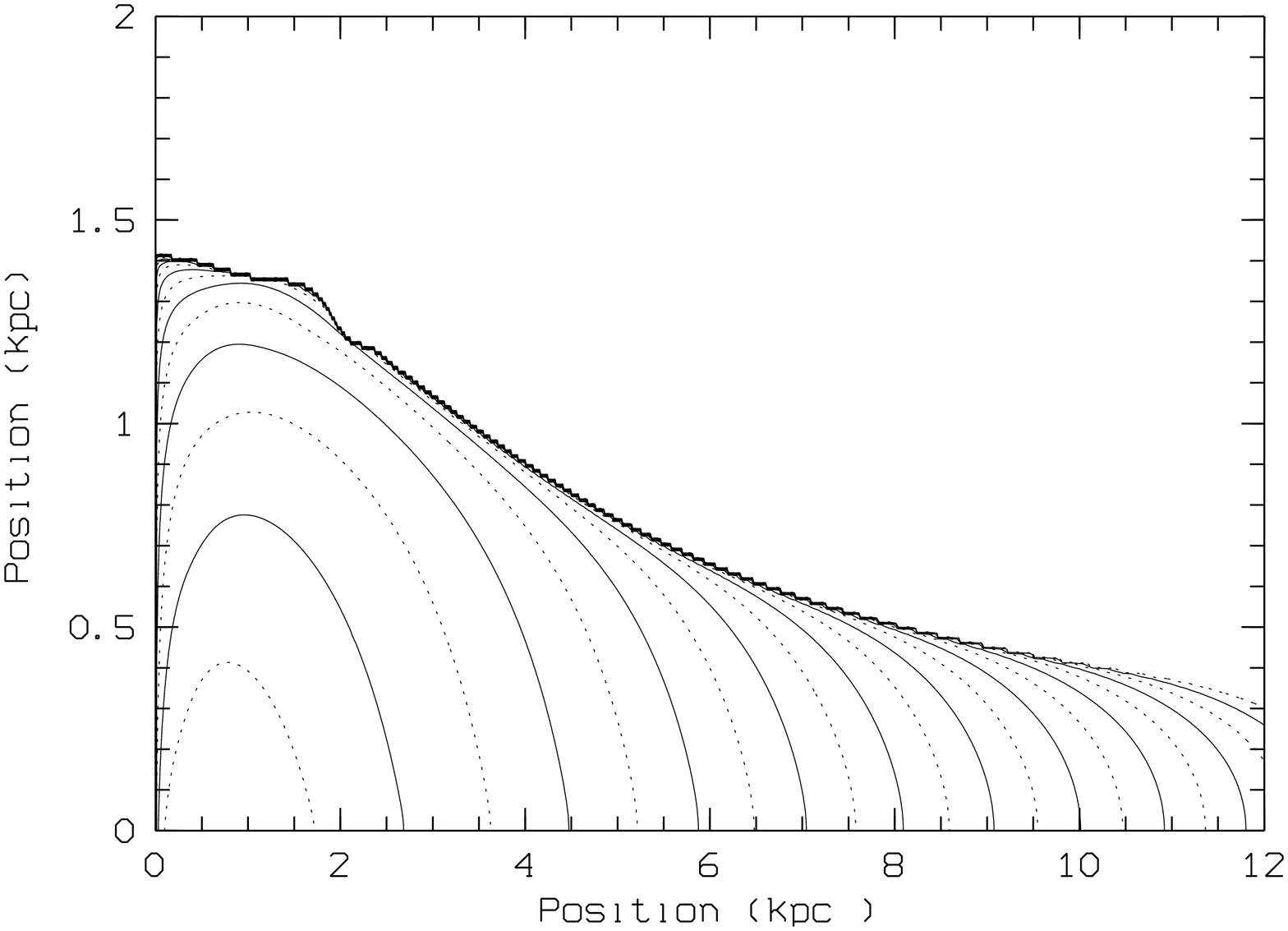,width=6cm,angle=-90}

\vbox{\vspace{-8.8cm}\hspace{8.8cm}
\psfig{figure=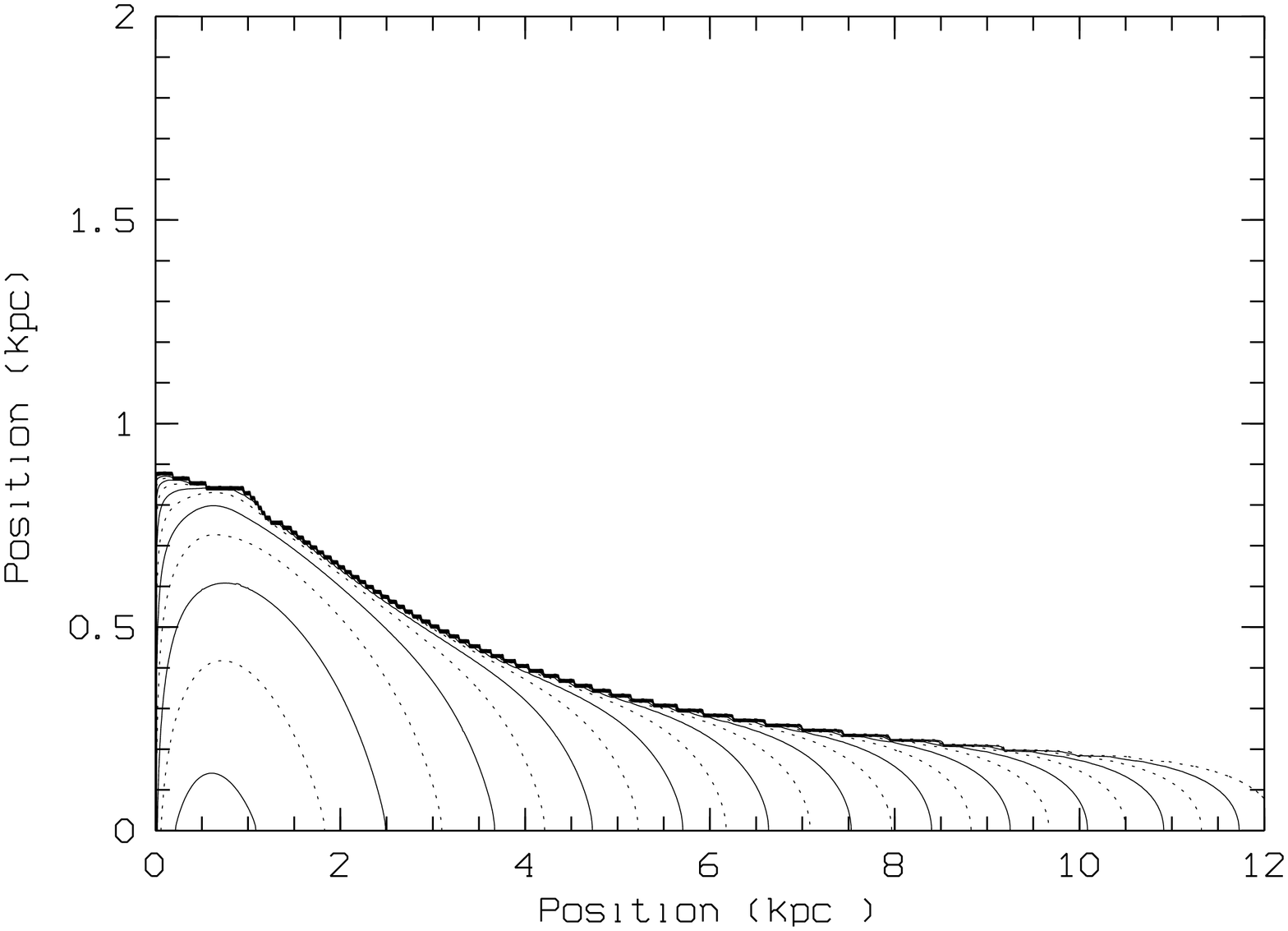,width=6cm,angle=-90}}

\caption{Contour plots of the mass density (i.e.\ the moment
$\mu_{0,0,0}$) in a meridional plane, for components with the
parameters $(\alpha_1,\alpha_2,\beta,\delta,\eta,z_0,s,a) =
(3,3,1,1,2,z_0,-0.5,-5)$, with $z_0$ equal to 4 kpc (left panel) and  2 kpc (right panel) respectively (note the very different scale for $\varpi$ and $z$). The discs become thinner with smaller
$z_0$, while the mass density is zero above $z=z_0$ (and even so below $z=z_0$ because $s \not= 0$). In this figure and in the following similar ones, every contour corresponds to a density that is a factor of 10 smaller than the next lower contour.}
\label{fig:parz0}
\end{figure}

This parameter was introduced in order to impose a maximum height above the galactic plane for the disc-like
component (Figure \ref{fig:parz0}): indeed, when $E \geq S_{z_0}$, an orbit cannot go higher than $z=z_0$, and the distribution function (\ref{eq:3icomp}) is null for $E \leq S_{z_0}(L_z)$. In order to model samples of stars belonging to populations
with different characteristic heights above the galactic plane, we can use a set of components with different values for this parameter.

\subsection{The parameter $\alpha_1$}

\begin{figure}
\psfig{figure=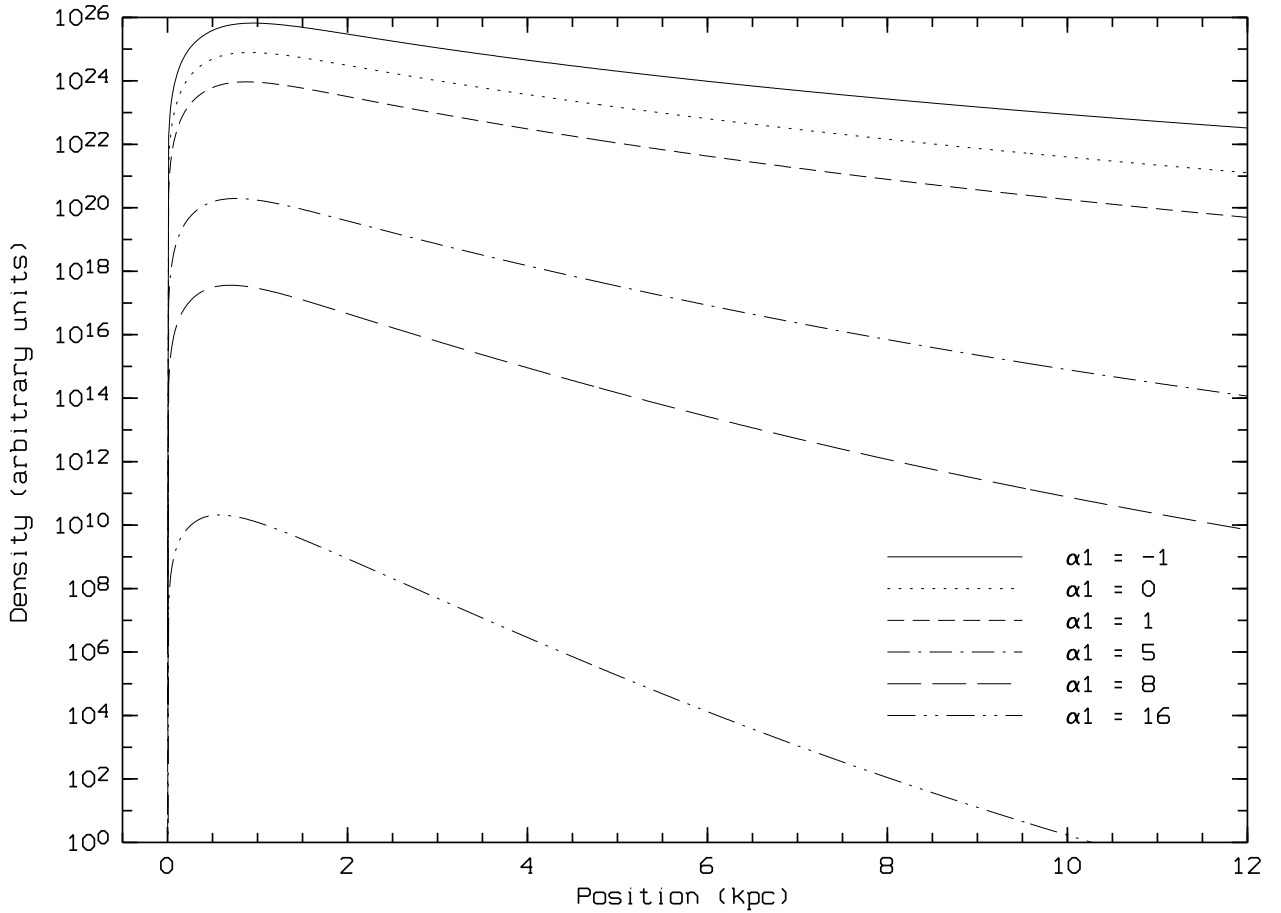,width=13.cm,angle=-90} \caption{This figure
 displays the decline of the logarithm of the galactic plane mass
 density (in arbitrary units and scaling) of different components for
 varying $\alpha_1$. A rising $\alpha_1$ helps to produce mass close
 to the center. The other parameters have the same values as in Figure
 4 (with $z_0$ equal to 2 kpc) except that $\alpha_2=0$.}
 \label{fig:a1}
\end{figure}

The parameter $\alpha_1$ enters Eq. (10) as the exponent of
$S_{z_0}$: so,  for non-negative values of $\alpha_1$,
the factor where it appears will behave as a declining function of
$L_z$, in the same way as $S_{z_0}(L_z)$ does, showing a steeper
decline for larger $\alpha_1$ (see Figure \ref{fig:a1}). A large $\alpha_1$ thus results in
a distribution function that favours a large fraction of bound
orbits. When it is increasing, this parameter helps to produce mass close to the center.

When a given exponential decline is requested, $\alpha_1$ will be a function of the other parameters rather than a fixed parameter (see section 5.3).

\subsection{The parameter $\alpha_2$}

\begin{figure}
\psfig{figure=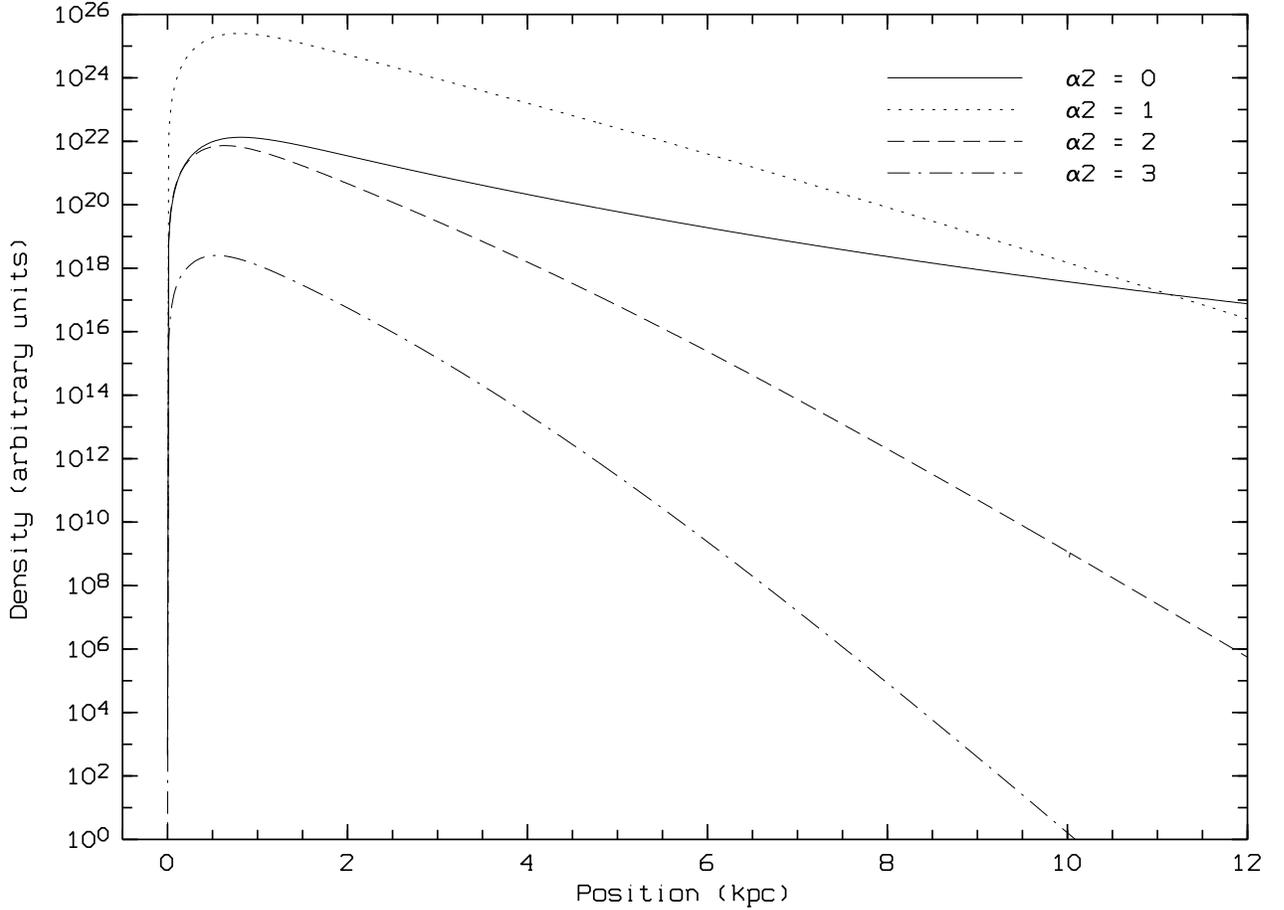,width=13.cm,angle=-90} \caption{Logarithm of the
 configuration space density (in the galactic plane and in arbitrary
 units) of different components for varying $\alpha_2$. The other
 parameters have the same values as in Figure 4 (with $z_0$ equal to 2
 kpc) except that $\alpha_1$ is adjusted to built-in a given exponential 
 decline. We see that the components have an exponential decline in the
 galactic plane and that $\alpha_2$ is roughly the reciprocal
 of the component's scale length.}  \label{fig:a2}
\end{figure}

The parameter $\alpha_2$ occurs as exponent in the distribution
function's exponential factor . Increasing values of this parameter
will contribute to the mass distribution near the center (Figure
\ref{fig:a2}), like in the $\alpha_1$ case (but exponentially).

On the other hand, our potential is approximately Keplerian at very
large radii: this implies that, in the galactic plane,
\begin{equation}
L_z^2 \sim \varpi
\end{equation}
and that (Batsleer \& Dejonghe 1995)
\begin{equation}
S_{z_0}(L_z) \sim \frac{1}{L_z^2} \sim \frac{1}{\varpi}
\end{equation}

So, at very large radii, $\alpha_2$ is the reciprocal of the
component's scale length, if the contribution of the other factors to
the mass density does not vary much with respect to $\varpi$ (for very
large $\varpi$). In practice, it is often desirable to use components
for which an exponential decline and a given scale length (as
determined by $\alpha_2$) is already built-in between two radii (say
$\varpi_1$ and $\varpi_2$). In such cases, the parameter $\alpha_1$
is adjusted in such a way that it corrects
for the non-constant behaviour of the other factors at large radii,
making the global contribution of all factors (except the one in
$\alpha_2$) constant at $\varpi_1$ and $\varpi_2$.

\subsection{The parameter $\beta$}

\begin{figure}
\psfig{figure=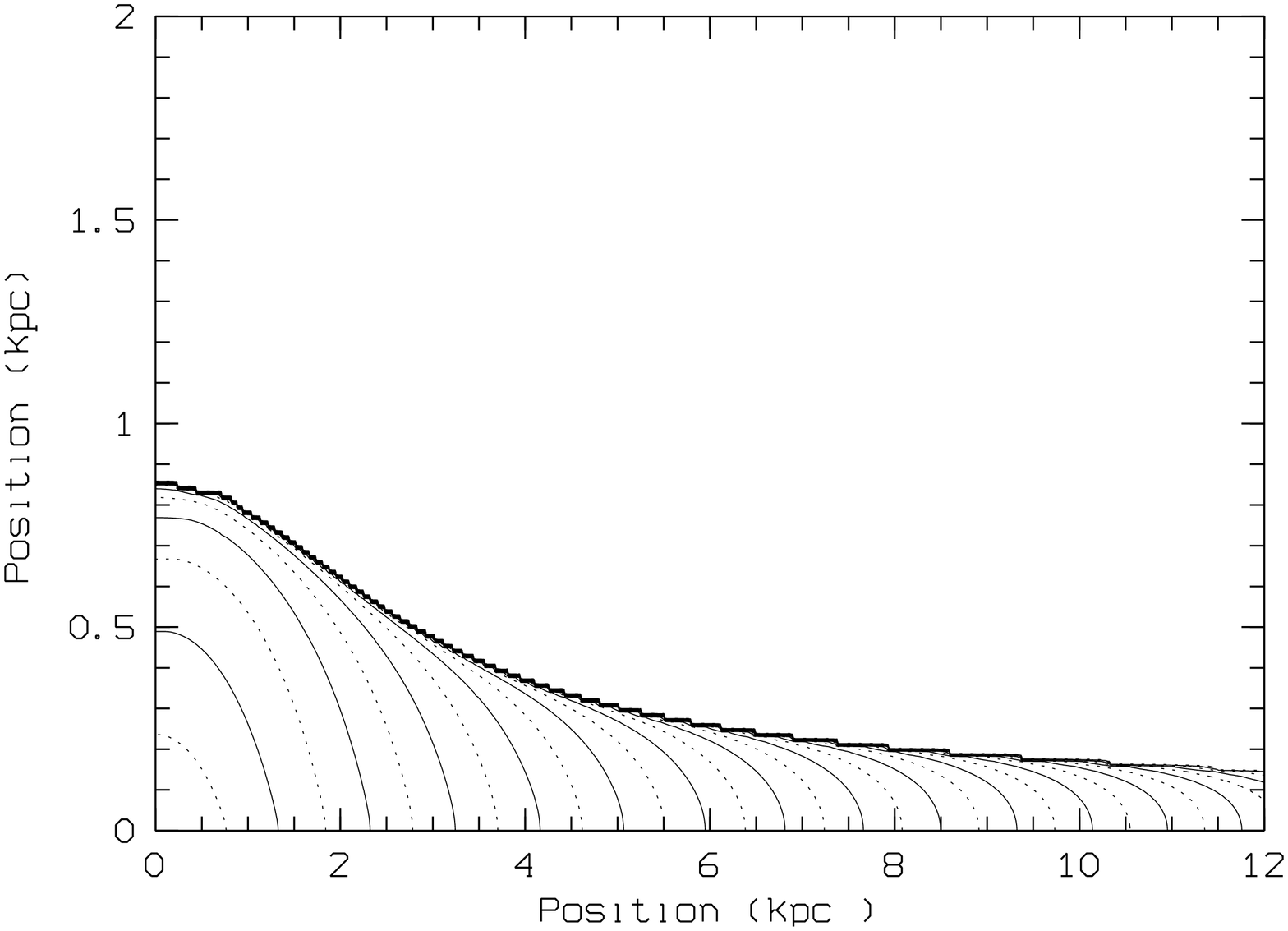,width=6cm,angle=-90}

\vbox{\vspace{-8.8cm}\hspace{8.8cm}
\psfig{figure=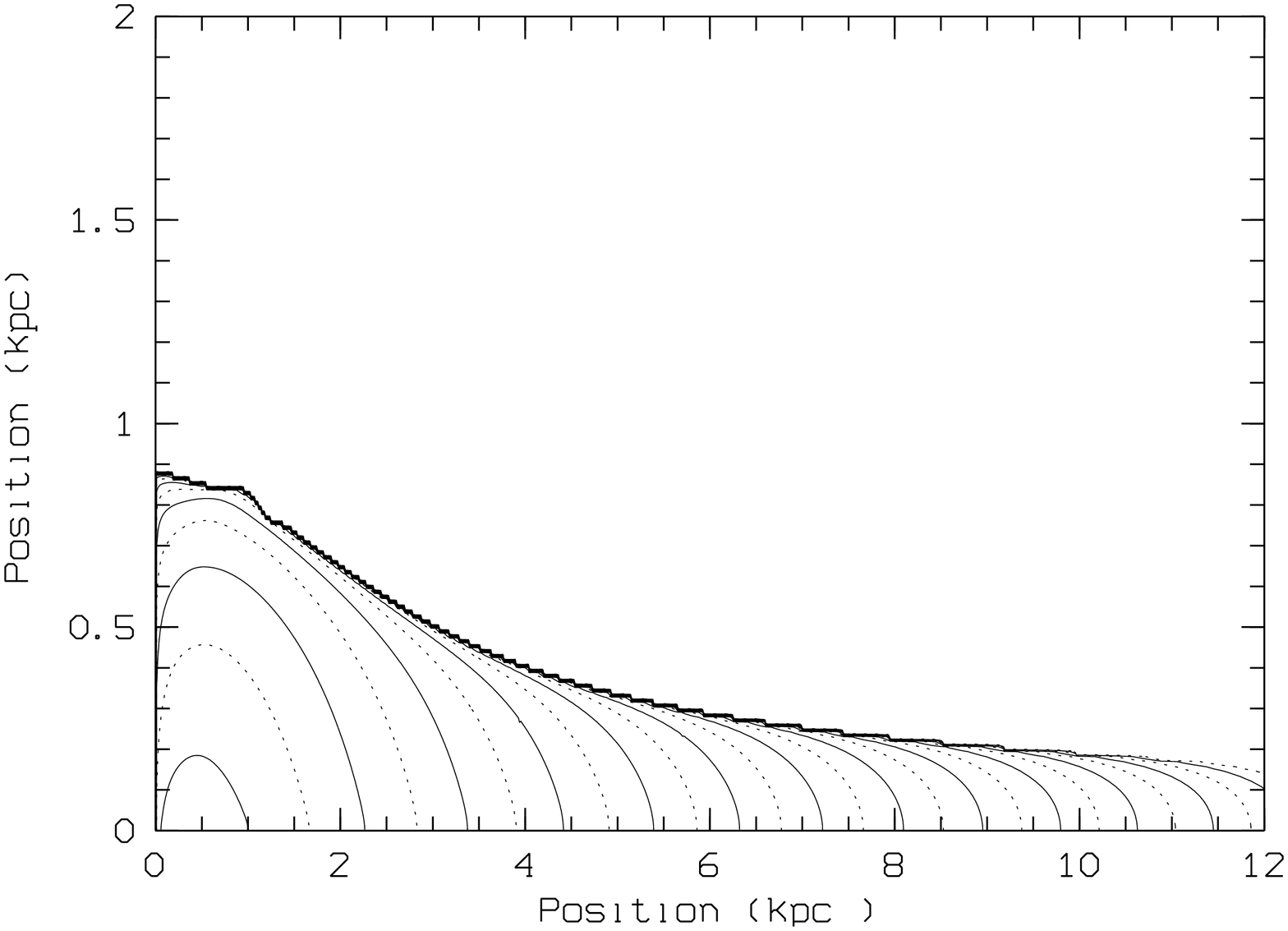,width=6cm,angle=-90}
}

\psfig{figure=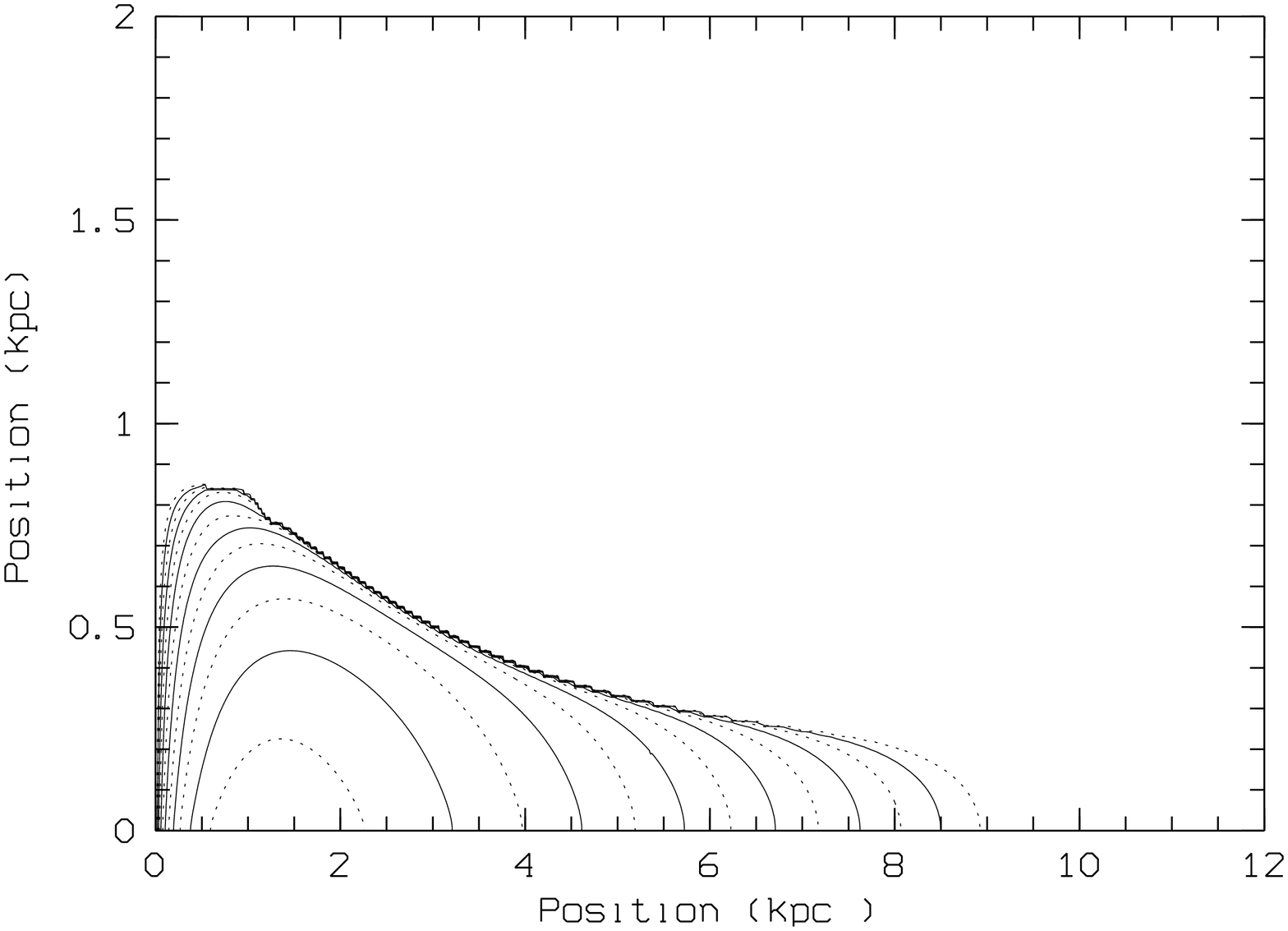,width=6cm,angle=-90}

\vbox{\vspace{-8.8cm}\hspace{8.8cm}
\psfig{figure=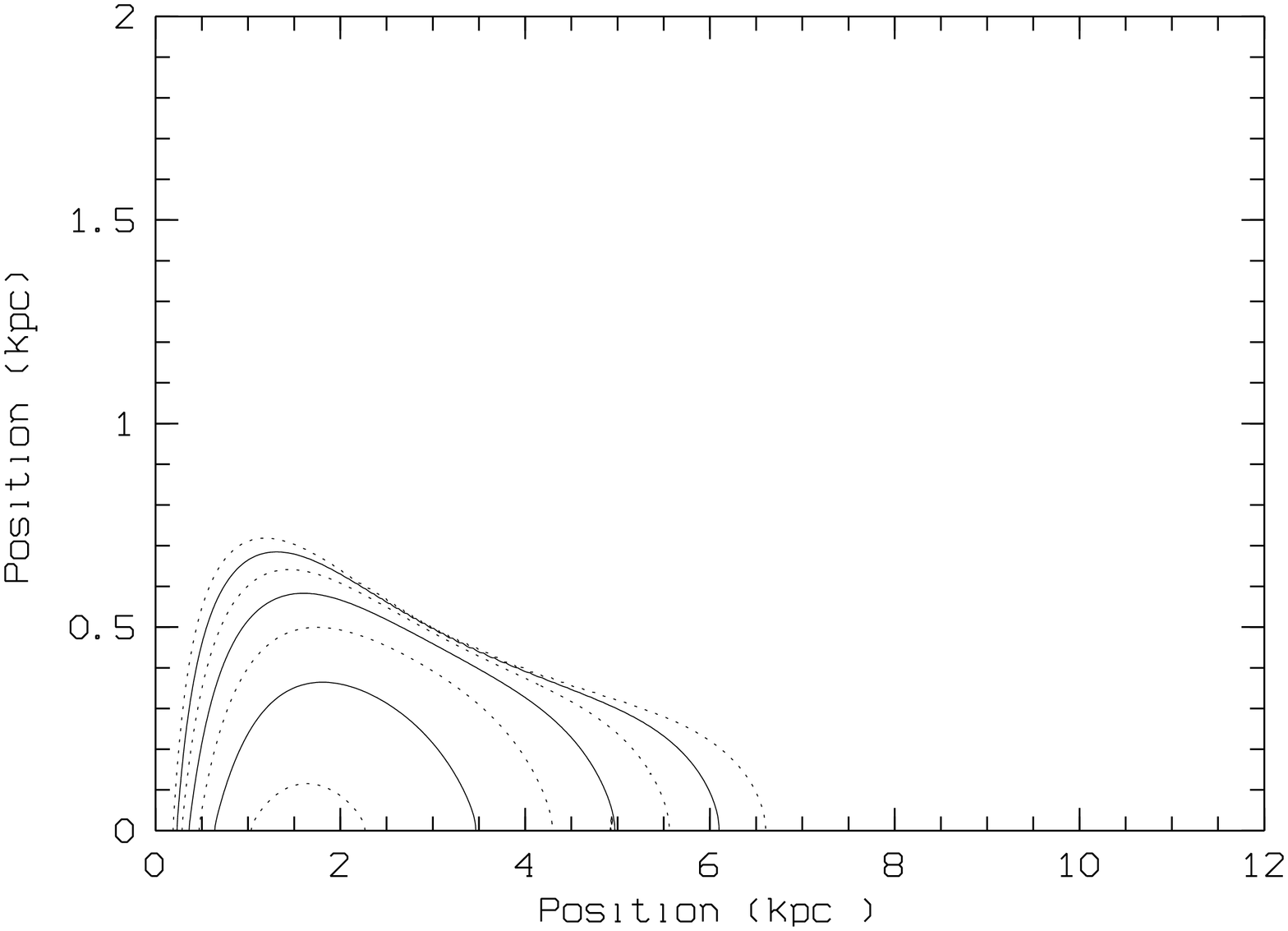,width=6cm,angle=-90}
}

\caption{Contour plots of the configuration space density (the mass density) in a meridional plane, for components with the
parameters $(\alpha_1,\alpha_2,\beta,\delta,\eta,z_0,s,a) =
(3,3,\beta,1,2,2,-0.5,-5)$, with $\beta=0$ (top left), $\beta=0.5$ (top right), $\beta=4$ (bottom left) and $\beta=6$ (bottom right). We see that the maximum number of stars moves away from the center and that the mass is more concentrated in configuration space for an increasing $\beta$.}
\label{fig:parb}
\end{figure}

For this parameter, there are two distinct cases: $\beta = 0$ and $\beta > 0$.
In the first case, the density is maximum in the center and falls off smoothly. In the latter case, the density is null in the center since $L_z = 0$ for $\varpi = 0$. In order to model real stellar systems, we need components with $\beta = 0$ to have some mass in the center. However, in a real galaxy, the maximum number of stars occurs in the intermediate region where the bulge meets the disc: this justifies the utilization of components with $\beta > 0$ when modelling real stellar systems. We see the maximum density moving away from the center when $\beta$ is rising (Figure \ref{fig:parb}). 

We also see on Figure (\ref{fig:parb}) that an increasing $\beta$ will concentrate the mass in a smaller region of configuration space.

\subsection{The parameter $\eta$}

\begin{figure}

\vbox{\epsfxsize=11cm \epsfbox{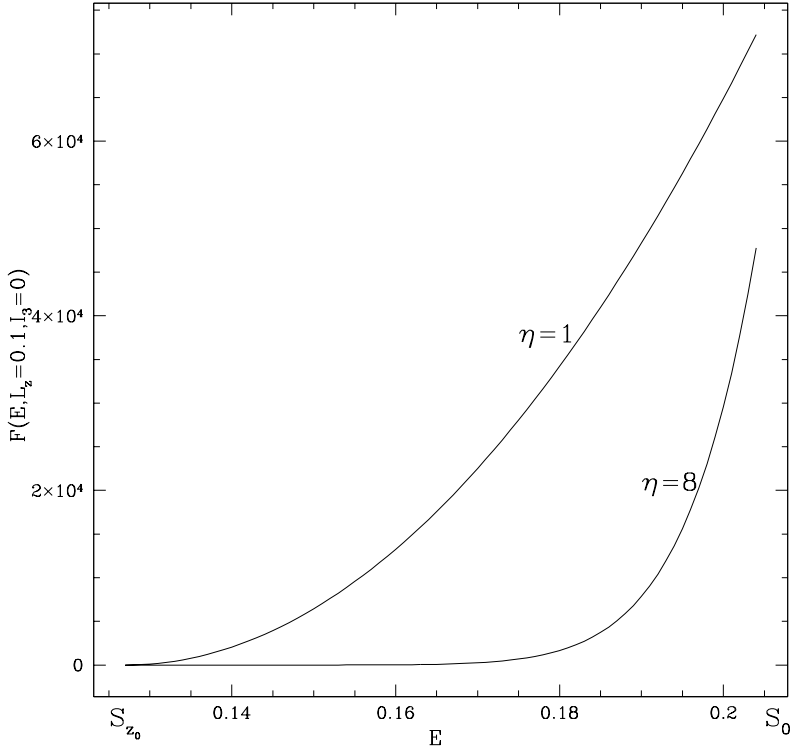}\vspace{-4cm}}
\hfill\parbox[b]{5.5cm}{\caption[]{For $I_3=0$ and $L_z=0.1$, this figure displays the value of a component distribution function (in arbitrary units) between $E=S_{z_0}$ and $E = S_0$ for $\eta=1$ and $\eta= 8$. The other parameters have the same values as in Figure 4 (with $z_0$ equal to 2 kpc). We see that the proportion of circular orbits (close to $E=S_0$) is much higher for $\eta=8$.}}  
  \label{fig:parg}
\end{figure}

For a given $L_z$, the largest value of the factor $(E-S_{z_0})^{\eta}$ is obtained when the binding energy $E = S_0$ corresponds to the circular orbits in the galactic plane (see Figure 1). 
So, the parameter $\eta$ is responsible for the favouring of almost circular orbits: a larger $\eta$ implies a larger contribution of almost circular orbits (Figure \ref{fig:parg}) and thus a mass density located closer to the plane.

\subsection{The parameter $s$}

\begin{figure}
\psfig{figure=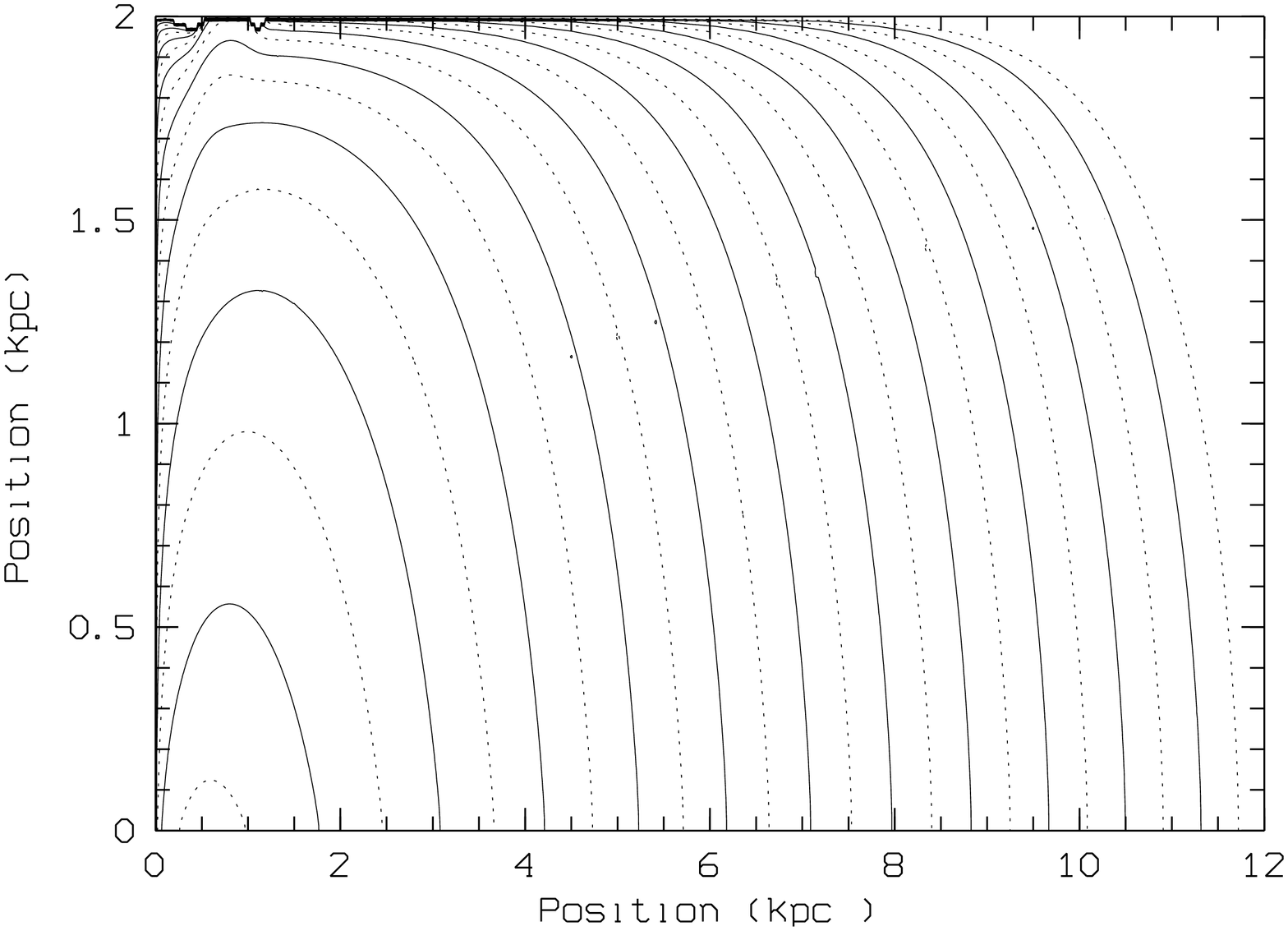,width=6cm,angle=-90}

\vbox{\vspace{-8.8cm}\hspace{8.8cm}
\psfig{figure=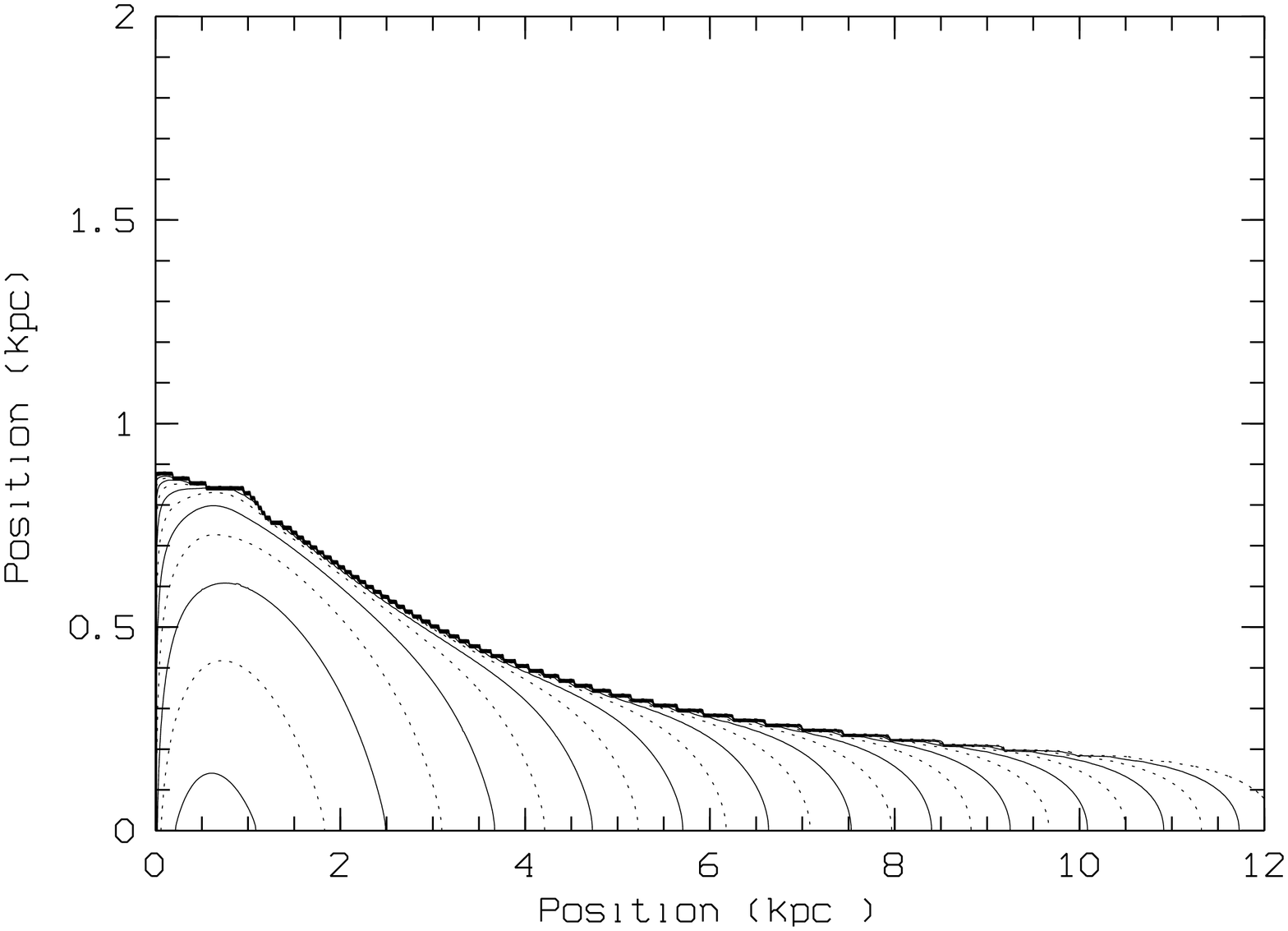,width=6cm,angle=-90}
}

\psfig{figure=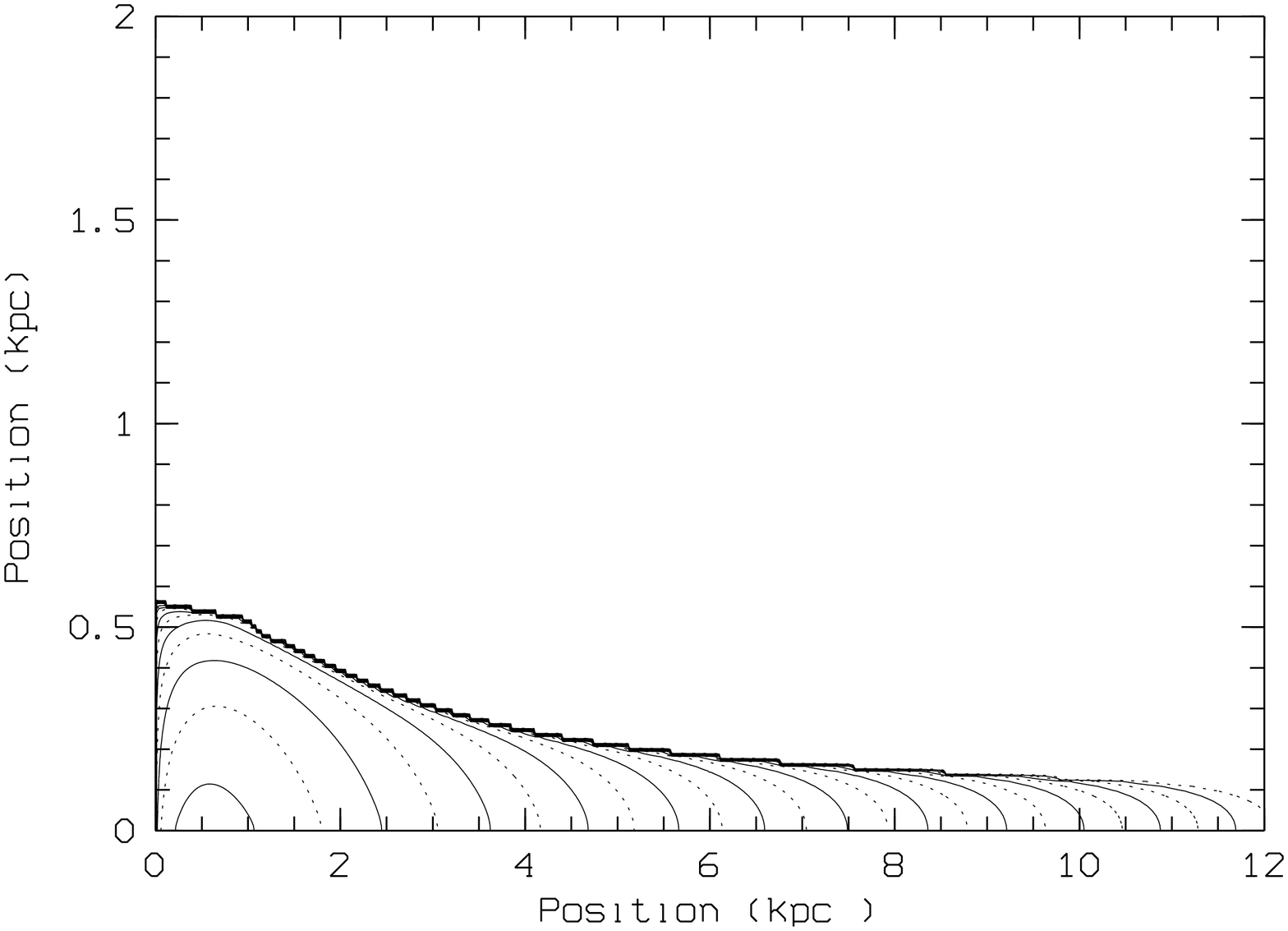,width=6cm,angle=-90}

\vbox{\vspace{-8.8cm}\hspace{8.8cm}
\psfig{figure=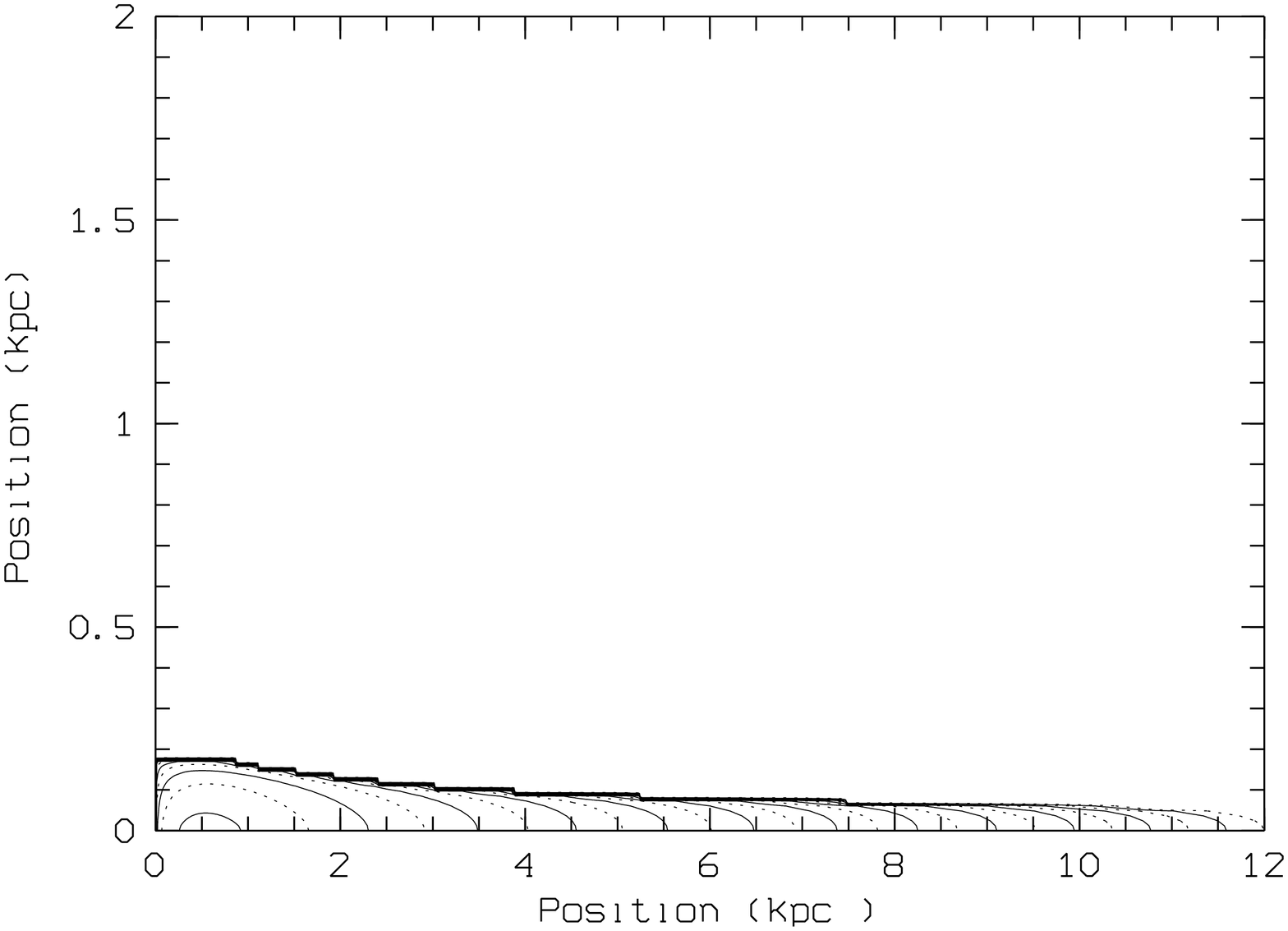,width=6cm,angle=-90}
}

\caption{Contour plots of the configuration space density (the mass density) in a meridional plane, for components with the
parameters $(\alpha_1,\alpha_2,\beta,\delta,\eta,z_0,s,a) =
(3,3,1,1,2,2,s,-5)$, with $s=0$ (top left), $s=-0.5$ (top right), $s=-1$ (bottom left) and $s=-4$ (bottom right). The height $z_0$ is reached only in the case $s=0$. The more negative $s$, the more the mass is concentrated near the galactic plane.}
\label{fig:pars}
\end{figure}

Condition (\ref{eq:cond})  $E \geq S_{z_0}-sI_3$ implies that for $I_3 \not= 0$ and a strictly negative $s$, the orbits cannot reach the height $z_0$ above the galactic plane. We see on Figure (\ref{fig:pars}) that the height $z_0$ is reached only in the case $s=0$. Furthermore, since a large $I_3$ corresponds to an orbit that can reach a large height, the factor $(E - S_{z_0} + sI_3)^{\delta}$ favours orbits that stay low. So, by setting $s$ more negative, we confine the orbits closer to the galactic plane. 

\begin{figure}
\psfig{figure=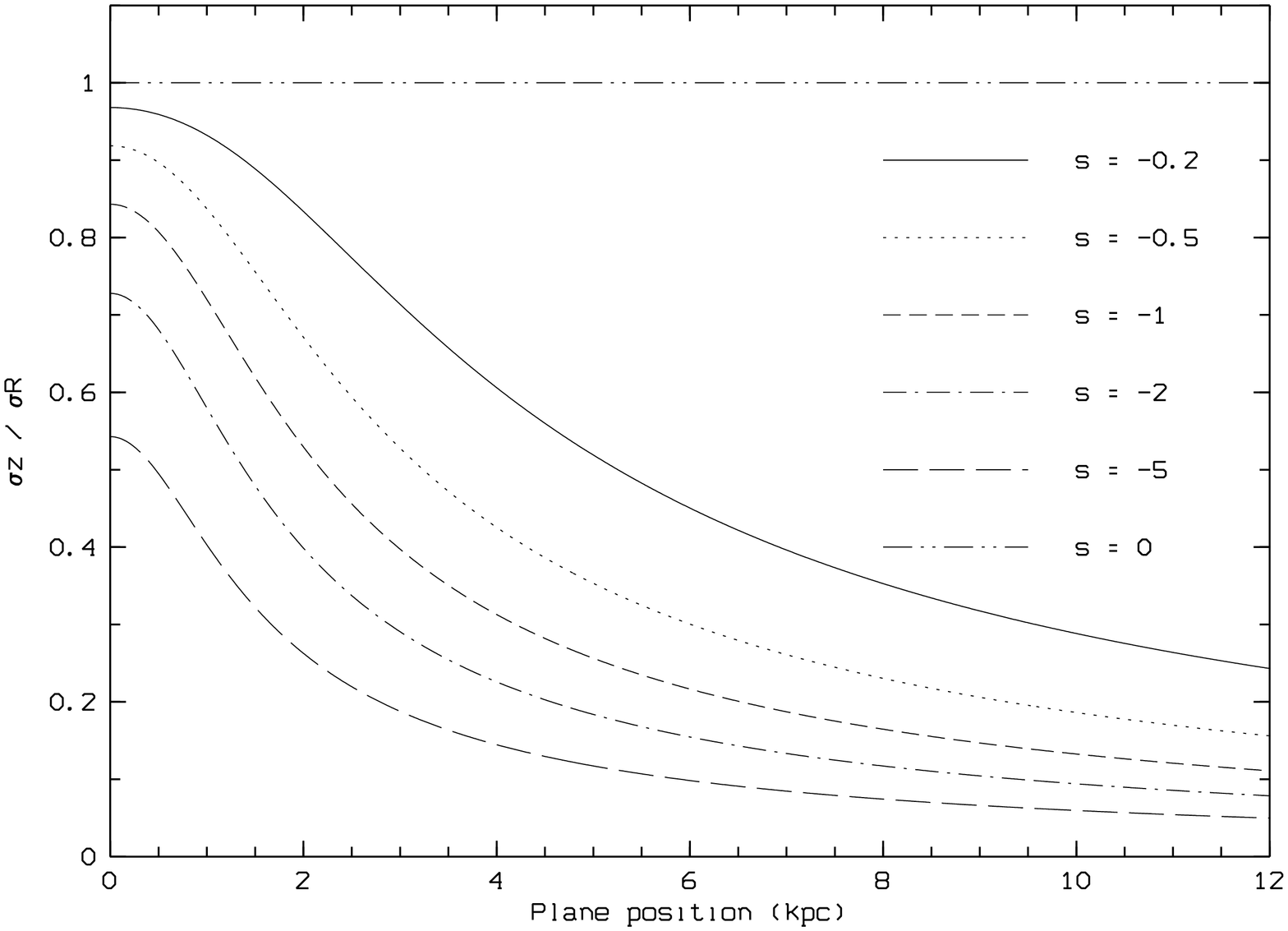,width=13.cm,angle=-90}
 \caption{This figure displays the ratio $\frac{\sigma_z}{\sigma_\varpi}$ of several components (in the Plane) for varying $s$. The other parameters have the same value as in Figure \ref{fig:pars}. The dependence of the components on the third integral induces anisotropy.}
 \label{fig:dispratio}
\end{figure}

A very important property of our components is the possibility of
introducing a certain amount of anisotropy in the stellar disc: if we
denote by $\sigma_z$ the dispersion of the velocity
in the direction perpendicular to the galactic plane, and by
$\sigma_\varpi$ the dispersion of the radial velocity in the galactic
plane, then any nonzero $s$ will produce a ratio
$\frac{\sigma_z}{\sigma_\varpi}$ less than 1 (Figure \ref{fig:dispratio}). The ratio is closer to unity in the center than in the outer regions: this indicates the physically realistic character of our components. For $s=0$, we find
$\sigma_z = \sigma_\varpi$ since we are dealing with a two-integral component again.

\subsection{The parameter $\delta$}

\begin{figure}
\psfig{figure=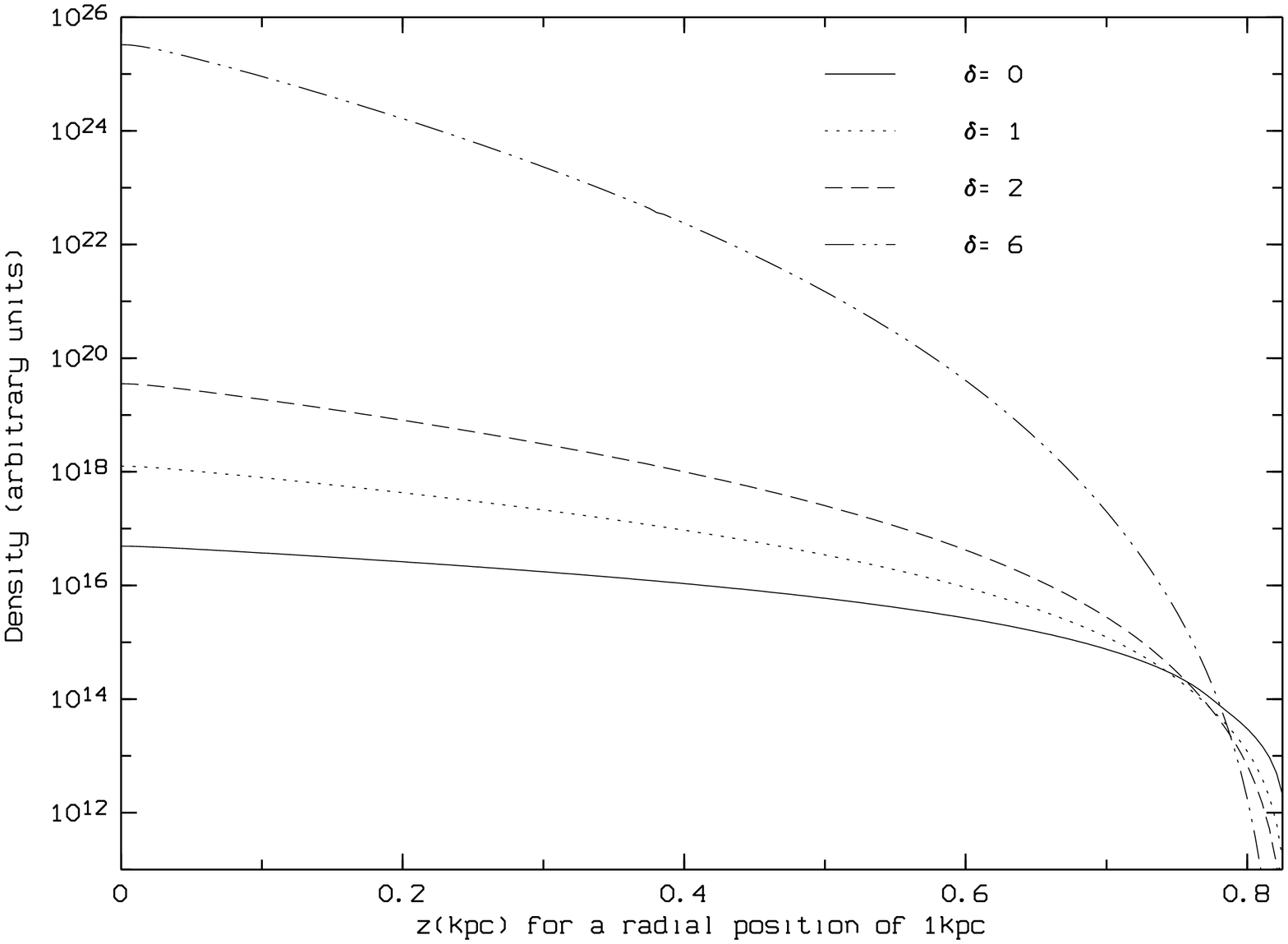,width=13.cm,angle=-90}
 \caption{Decline of the logarithm of the configuration space density (in arbitrary units) as a function of the height above the Galactic plane at $\varpi=1$ kpc for varying $\delta$. A rising $\delta$ implies a steeper decline. The other parameters have the same values as in Figure 4 (with $z_0$ equal to 2 kpc).}
 \label{fig:pard}
\end{figure}

A large $\delta$ has partly the same effects as a large $\eta$: it favours circular orbits. Furthermore, a large $\delta$ augments the effects of the negative $s$ and forces the stars to stay close to the plane by favouring low $I_3$-values. As we can see on Figure (\ref{fig:pard}), a component with a larger $\delta$ has more stars in the galactic plane but shows a steeper decline with respect to $z$.

\section{Modelling}

\begin{figure}

\psfig{figure=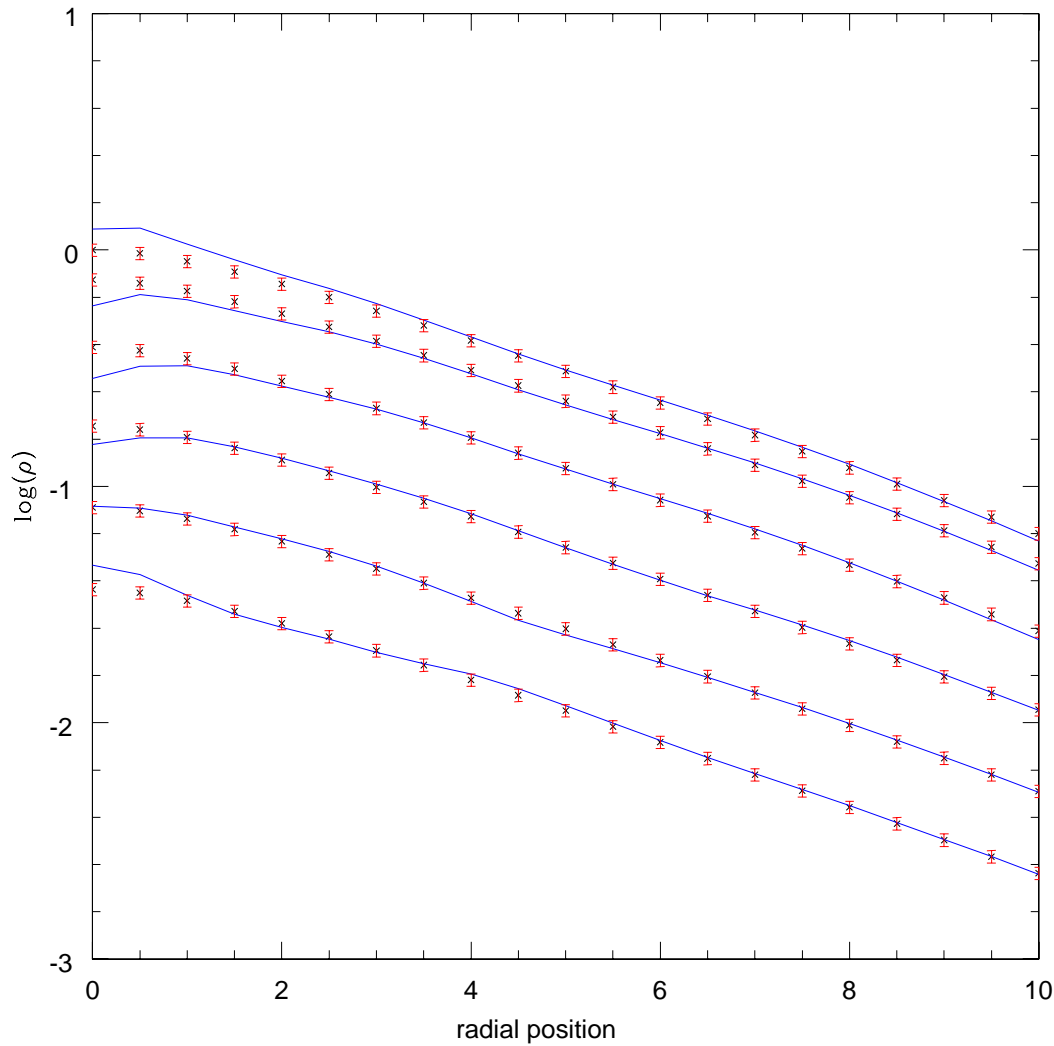,width=15cm,angle=0}
\caption{Data for a van der Kruit disc ($h_R = 3$kpc and $h_z = 0.25$kpc) were
generated in the region $0 \mathrm{ ~kpc} \leq \varpi \leq 10 \mathrm{ ~kpc}$
and $0 \mathrm{ ~kpc} \leq z \leq 1 \mathrm{ ~kpc}$. The initial subset of components was  made of the components with $\beta=0,1,3,5,7$;  $\alpha_1=1$;  $\alpha_2=0.15,0.3,2$;  $z_0=1,2,4$;  $\eta=1,5,10$;  $s=0,-0.5,-1$;  $\delta=0.01,1,4$;  $a=0$. Components with non-zero $s$ and non-zero $\delta$ were selected by the QP program. The crosses indicate the data for $z=0$pc, $200$pc, $400$pc, $600$pc, $800$pc, $1$kpc (from top to bottom), with error bars. The solid lines correspond to the mass density of the components linear combination at these heights. The only region of our disc where the fit deviates a little from the data is the $2$ kpc central region: our components are primarily intended to describe the outer regions rather than the central region of galaxies since most of them have a supermassive black hole at the center, sometimes associated with deviations from axisymmetry (bar).}  
  \label{fig:fit}
\end{figure}

\begin{figure}
\psfig{figure=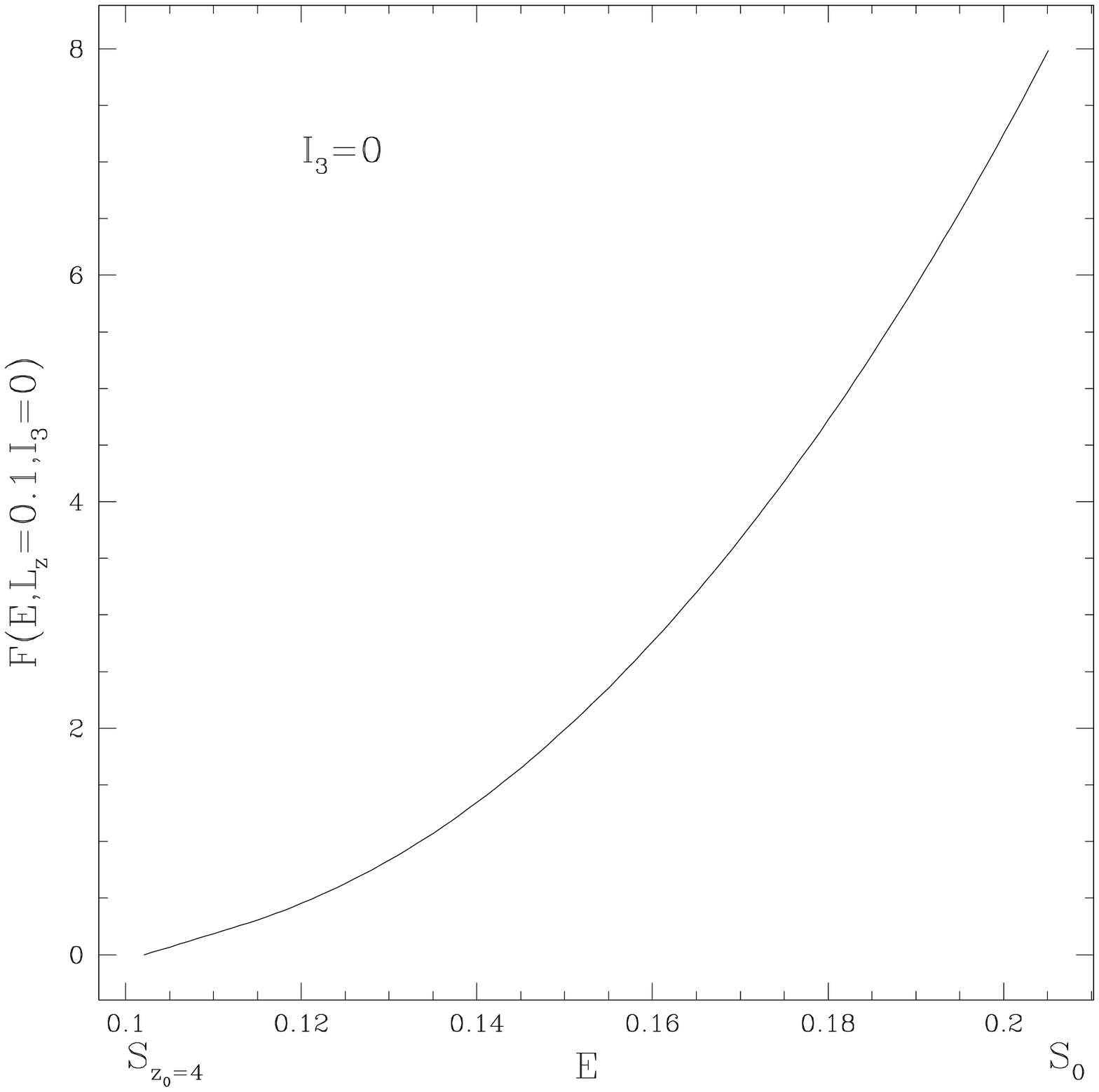,width=8.8cm}

\vbox{\vspace{-8.8cm}\hspace{8.8cm}
\psfig{figure=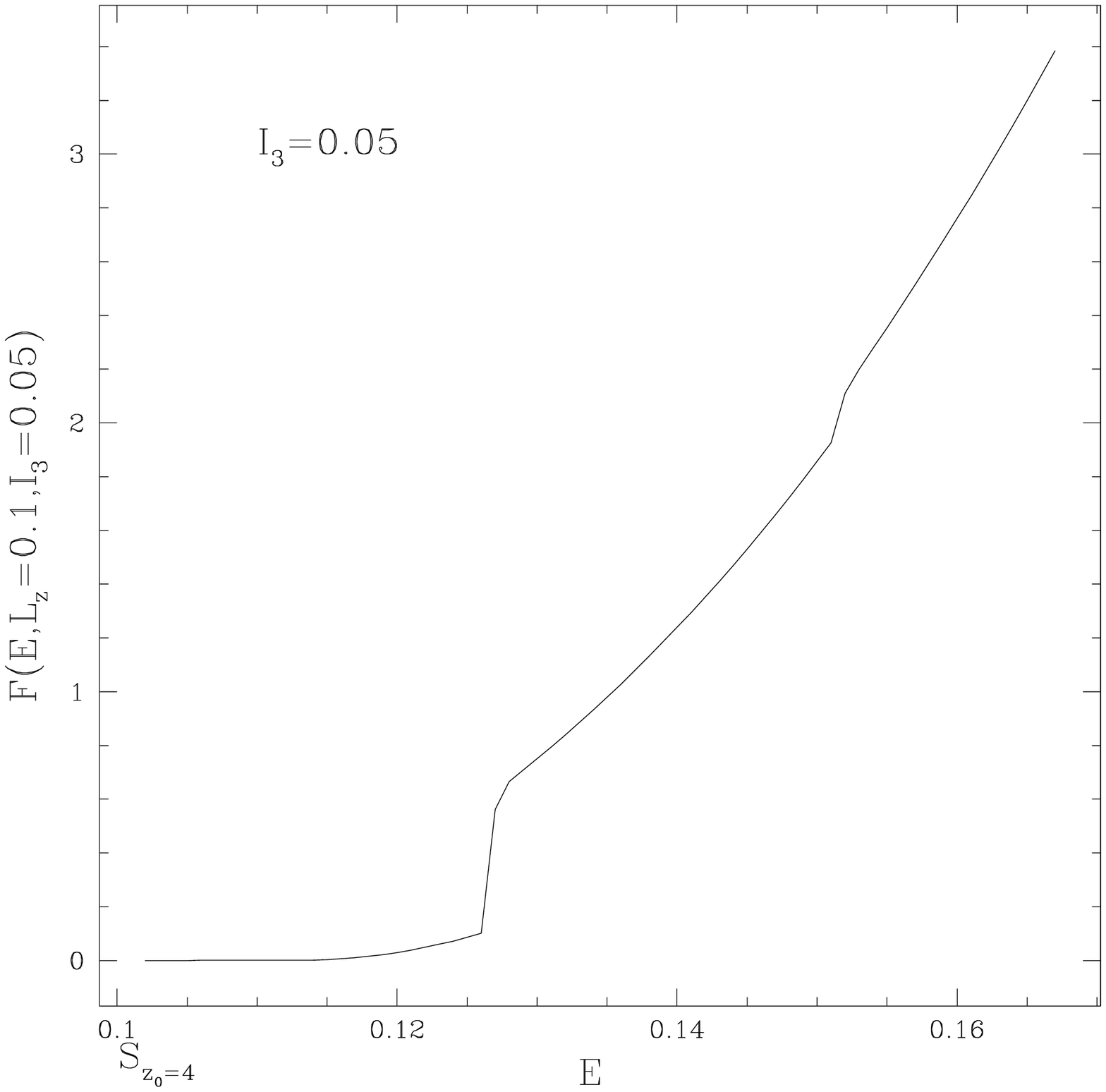,width=8.8cm}}

\caption{For $L_z=0.1$ and a fixed $I_3$ (left panel:$I_3=0$, riht panel:$I_3=0.05$), this figure displays the values of the distribution function (corresponding to the fit obtained in Figure 12) as a function of $E$ (for the bound orbits). For $I_3=0.05$, the maximum value of $E$ is the one corresponding to thin tube orbits and is smaller than $S_0$.}
\end{figure}

\begin{figure}
\psfig{figure=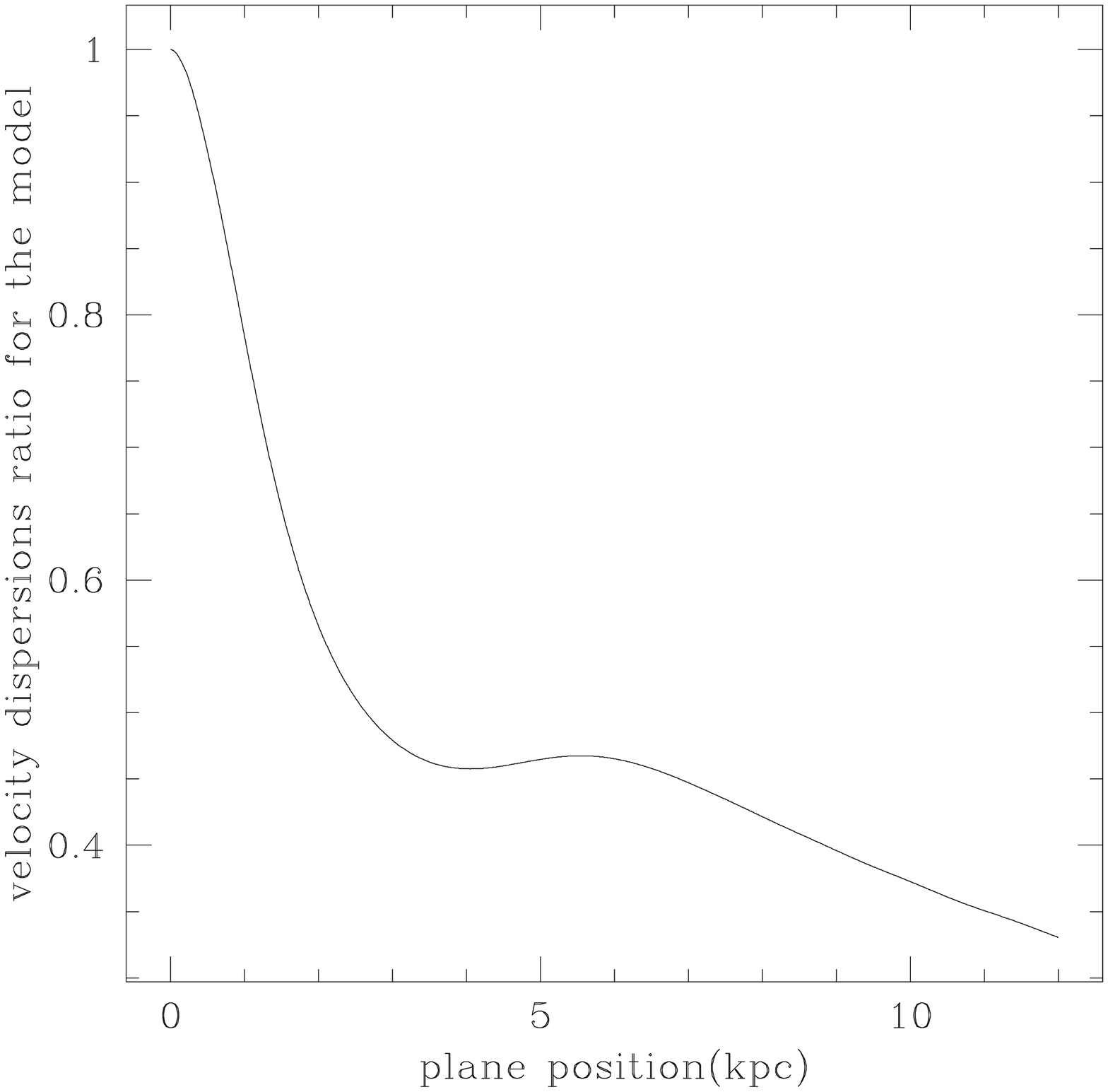,width=8.8cm}

\vbox{\vspace{-8.8cm}\hspace{8.8cm}
\psfig{figure=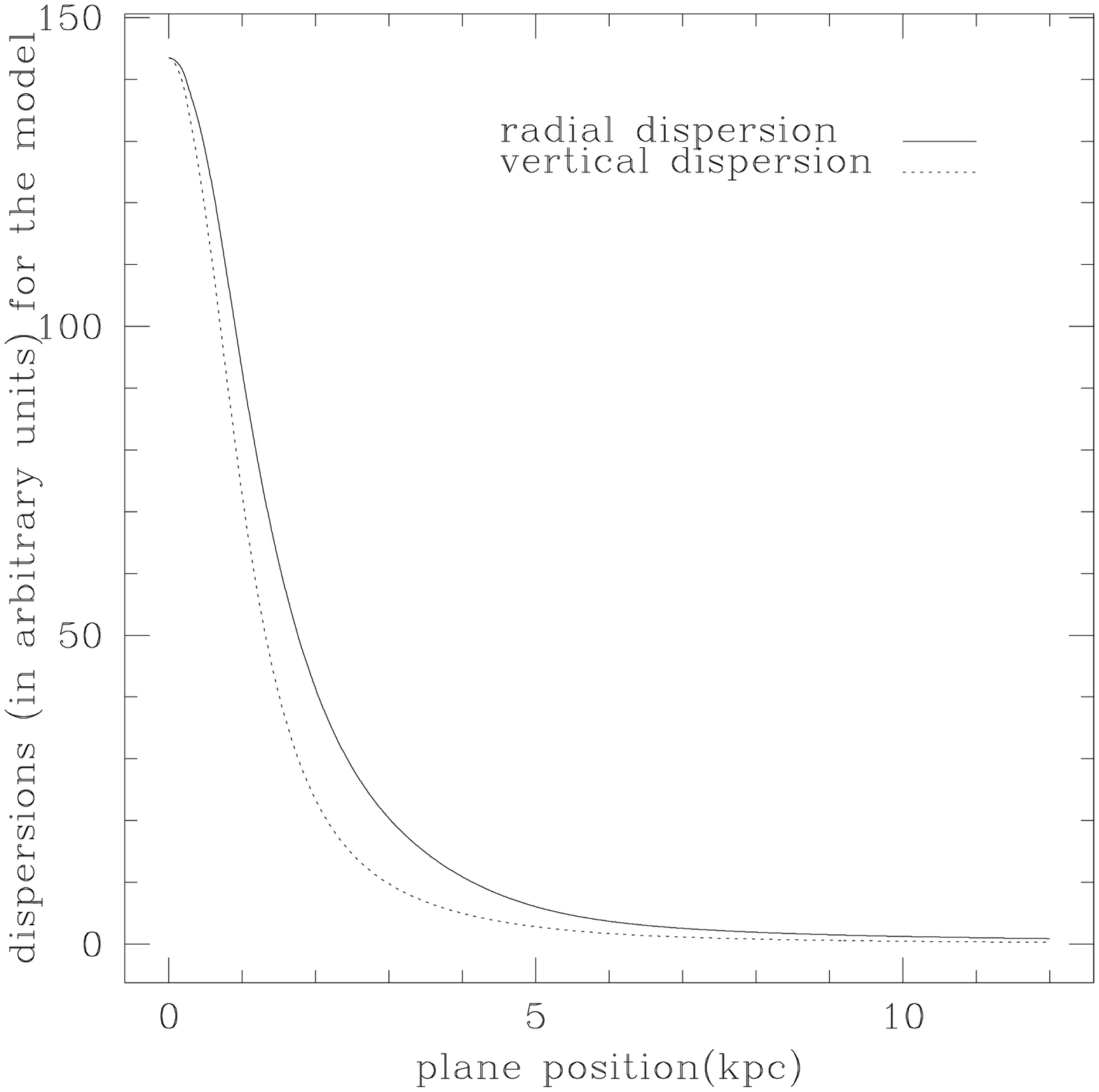,width=8.8cm}}

\caption{The left panel displays the ratio $\frac{\sigma_z}{\sigma_\varpi}$ in the galactic plane for the fit obtained in Figure 12. The right panel gives the shapes of the individual velocity dispersions curve $\sigma_\varpi$ (solid line) and $\sigma_z$ (dotted line) in the galactic plane.}
\end{figure}

The component distribution functions described in this paper are very
useful as basis functions in the method described by Dejonghe\ (1989),
in order to model any observable quantities (spatial mass density,
velocity dispersions, average radial velocities on a sky grid,...). As
an illustration, we present the application of the method to fit a
given spatial density $\rho_0(\varpi,z)$ (see Batsleer \& Dejonghe
1995 for a similar application in the two-integral approximation).  We
look for a linear combination of our components
\begin{equation}
 \sum_{\Lambda} c_{\Lambda}F_{\Lambda}
\end{equation}
that fits $\rho_0(\varpi,z)$, with $\Lambda =
(\alpha_1,\alpha_2,\beta,\delta,\eta,z_0,s,a)$ and $c_\Lambda$ the
coefficients that are to be determined.

In practice, to find this linear combination we must introduce a grid $(\varpi_i,z_i)$ in configuration space  and minimize the quadratic function in $c_{\Lambda}$:
\begin{equation}
\chi^2 = \sum_{i} \left[\left(\sum_{\Lambda} c_{\Lambda}\mu_{0,0,0}^{(\Lambda)}(\varpi_i,z_i) \right) - \rho_0(\varpi_i,z_i) \right]^2
\label{eq:chi2}
\end{equation}
 \noindent
This minimization, together with the constraint that the distribution function must be positive in phase space, is a problem of quadratic programming (hereafter QP) described by Dejonghe\ (1989).

Here, we choose to adopt for $\rho_0$ a spatial density which closely resembles that of a real disc, i.e. a van der Kruit law, for which the vertical disribution is a good compromise between an exponential and an isothermal sheet\ (van der Kruit 1988).
\def\sech{\mathop{\rm sech}\nolimits}
\begin{equation}
        \rho_0(\varpi,z) \propto \exp \left(-\frac{\varpi}{h_R} \right) 
                \sech \left( \frac{z}{h_z} \right)
\end{equation}
\noindent 
In order to have a zero derivative with respect to $\varpi$ on the
rotation axis, we adopt a mass density that follows closely the van
der Kruit law, without a cusp in the center (see also Batsleer \&
Dejonghe\ 1995):

\begin{equation}
        \rho_0(\varpi,z) = \frac{1+2 \varpi / h_R}{1+ \varpi / h_R}
                \exp \left(-\frac{\varpi}{h_R} \right) 
                \sech \left( \frac{z}{h_z} \right),
\end{equation}
\noindent
with $h_R$ and $h_z$ denoting the horizontal and vertical scale
factor, respectively.

Since the moments $\mu_{0,0,0}$ are dependent on the potential
of the galaxy (including the dark matter),
we have to choose a potential for the galaxy that contains the stellar
disc we want to model. We adopt a St\"ackel potential with three mass
components that produces a flat rotation curve and that therefore is a
candidate potential for a disc galaxy (Famaey \& Dejonghe 2001, see
also Batsleer \& Dejonghe 1994).

The actual modelling follows the same strategy as followed by
Batsleer \& Dejonghe\ (1995), which we briefly repeat here for easy
reference. 
The first step in the actual modelling consists in the selection of a
subset of components out of the (infinite) set of possible components.
This subset is chosen so that certain features, that we suppose to be
present in the stellar disc, such as circular orbits, are
included. For example, we expect the mass density corresponding to a
component to have an exponential behaviour close to the mass density
we want to model. The QP program first minimizes the function
(\ref{eq:chi2}) for one component $F_{\Lambda}$ and chooses the
component of the initial subset that produces the lowest minimum for
that function (\ref{eq:chi2}).  Then the program iterates, selecting
and adding at each iteration the component which, together with the
components already chosen in a previous run, produces the best fit.
Once the minimum of the $\chi^2$-variable does not change
significantly any more with the addition of extra components, the
program is halted because too low a value for $\chi^2$ could imply
that the QP program starts producing a distribution function featuring
unnecessary oscillations.

As an example, we model a modified van der Kruit disc with $h_R = 3$kpc and
$h_z = 0.25$kpc. Batsleer \& Dejonghe\ (1995) already showed that a
linear combination of two-integral components (with $s=0$ and
$\delta=0$) could fit such a disc, but with $\sigma_\varpi =
\sigma_z$. In order to model real anisotropic velocity data in the
future, the dependence on the third integral will be needed. We show
that, by choosing components with $\beta=0,1,3,5,7$; $\alpha_1=1$;
$\alpha_2=0.15,0.3,2$; $z_0=1,2,4$; $\eta=1,5,10$; $s=0,-0.5,-1$;
$\delta=0.01,1,4$ and $a=0$ in the initial subset, a fit with
components featuring $s \not= 0$ and $\delta \not= 0$ can be obtained
too (see Figure 12).

The fit is obtained for a linear combination of
$25$ components at $231$ configuration space points ($206$ degrees of
freedom). If we assume relative errors of $6 \%$, we obtain for our
minimum $\chi^2=246$, and the probability that a value of $\chi^2$
larger than $246$ should occur by chance is $Q(246/2,206/2) \simeq
0.1$\ (Abramowitz \& Stegun 1972), which makes the goodness-of-fit
believable\ (Press et al. 1986).

By using St\"ackel dynamics to model a galactic disc, we construct a
completely explicit and analytic distribution function, with an
explicit dependence on the third integral. Figure (13) displays the
distribution function obtained by QP in function of $E$, for $L_z=0.1$
and for two values of $I_3$ ($I_3=0$ and $I_3=0.05$). For $I_3=0$, the
distribution function is non-zero if $S_{z_0}\leq E \leq S_0$ (with
$z_0=4$kpc); for $I_3 > 0$, instead, the maximum value of $E$ is the
one corresponding to infinitesimally thin short axis tubes and is
smaller than $S_0$. We see on Figure (13) that the distribution
function is decreasing with increasing $I_3$ (particularly near
$E=S_{z_0}$), and that it has some clumps. These clumps at $I_3=0.05$
are not discontinuities since the distribution function is a linear
combination of continuous components.

Many different three-integral distribution functions correspond to a
given spatial density, and there is no guarantee that they will yield
realistic velocity dispersions. It is a major result of this paper to
show that it is possible to find a linear combination of our
components yielding realistic velocity dispersions. Figure (14)
displays the ratio $\frac{\sigma_z}{\sigma_\varpi}$ in the galactic
plane: at the radius corresponding to the solar position in the Milky Way ($7.5$-$8.5$ kpc),
the classical value of $\frac{\sigma_z}{\sigma_\varpi} \simeq 0.4$ is
obtained. The local maximum in the $\frac{\sigma_z}{\sigma_\varpi}$
curve is due to the individual shapes of the velocity dispersions
curves (Figure 14).

\section{Conclusions}

In this paper, we have constructed new analytic three-integral stellar distribution functions yielding $\sigma_{\varpi} \not= \sigma_z$: they are generalizations of two-integral ones that can describe thin discs with the restriction that $\sigma_{\varpi}=\sigma_z$\ (Batsleer \& Dejonghe 1995). 

We first reduced the triple integral defining their moments to a simple one, like in the Abel case\ (Dejonghe \& Laurent 1991), by making some assumptions on the parameters. Then we looked for the effects of the different parameters and showed the disc-like (physically realistic) features of our distribution functions: they have a finite extent in vertical direction and an exponential decline in the galactic plane, while favouring almost circular orbits. A very important feature induced by the dependence on the third integral is their ability to introduce a certain amount of anisotropy, by varying the parameters responsible for this dependence ($s$ and $\delta$).

We  finally showed that a van der Kruit disc can be modelled by a linear combination of such distribution functions with an explicit dependence on the third integral and a realistic anisotropy in velocity dispersions. This implies that they are very promising tools to model real data with $\sigma_{\varpi} \not= \sigma_z$ (Hipparcos data for example) by using the quadratic programming algorithm described by Dejonghe\ (1989). This will provide information on the dynamical state of tracer stars in the Milky Way (or on external galaxies).

\section*{Acknowledgements}

We thank Dr Alain Jorissen very much for his permanent assistance. We thank the referee Dr Stephen Levine for his thorough reading of the manuscript and many helpful suggestions.

\label{lastpage}

\end{document}